\documentclass[10pt,peerreview]{IEEEtran}

\makeatletter
\def\ps@headings{%
\def\@oddhead{\mbox{}\scriptsize\rightmark \hfil \thepage}%
\def\@evenhead{\scriptsize\thepage \hfil\leftmark\mbox{}}%
\def\@oddfoot{}%
\def\@evenfoot{}}
\makeatother
\usepackage{cite}
\usepackage{amsmath}
\usepackage{amssymb}
\usepackage{subfigure}
\usepackage{color}
\usepackage{times}
\usepackage{url}
\usepackage{graphicx}
\usepackage{epstopdf}
\usepackage{amsthm}  

\usepackage{tabularx}

\usepackage{soul}

\newtheorem{theorem}{Theorem}
\newtheorem{definition}{Definition}
\newtheorem{lemma}[theorem]{Lemma}
\newtheorem{example}{Example}

\newtheorem{prop}[theorem]{Proposition}
\newtheorem{property}{Property}

\newtheorem{construction}{Construction}
\usepackage[ruled,noline]{algorithm2e}

\newcommand{\parent}[1]{\left( #1 \right)}

\newcommand{\be}{\begin{equation}}
\newcommand{\ee}{\end{equation}}
\newcommand{\bp}{\begin{proof}}
\newcommand{\bpo}{ \begin{proof}[Proof Outline] }
\newcommand{\ep}{\end{proof}}

\newcommand{\Prob}[1]{\Pr\left( #1 \right)}

\newcommand{\PInv}{{Padding-Invariant }}
\newcommand{\pinv}{{padding-invariant }}

\newcommand{\ArcName}{memory }

\newcommand{\ftag}{G}

\newcommand{\graphheightA}{0.23}  

\newcommand{\Conf}[1]{}
\newcommand{\ExtendedJournal}[1]{}
\newcommand{\Journal}[1]{{#1}}

\newcommand{\OldText}[1]{}

\newcommand{\SingleColMode}[1]{{#1}}
\newcommand{\TwoColMode}[1]{}

\newcommand{\kl}{k_1}
\newcommand{\kh}{k_2}
\newcommand{\dc}{d}
\newcommand{\len}{l}

\newcounter{commentNumberI}

\begin{document}

\title{Optimal Compression for Two-Field Entries in Fixed-Width Memories}

\author{\IEEEauthorblockN{\large Ori Rottenstreich and Yuval Cassuto}}

\maketitle


\begin{abstract}
Data compression is a well-studied (and well-solved) problem in the setup of long coding blocks. But important emerging applications need to compress data to memory words of small fixed widths. This new setup is the subject of this paper. In the problem we consider we have two sources with known discrete distributions, and we wish to find codes that maximize the success probability that the two source outputs are represented in $L$ bits or less. A good practical use for this problem is a table with two-field entries that is stored in a memory of a fixed width $L$. Such tables of very large sizes drive the core functionalities of network switches and routers. After defining the problem formally, we solve it optimally with an efficient code-design algorithm. We also solve the problem in the more constrained case where a single code is used in both fields (to save space for storing code dictionaries). For both code-design problems we find decompositions that yield efficient dynamic-programming algorithms. With an empirical study we show the success probabilities of the optimal codes for different distributions and memory widths. In particular, the study demonstrates the superiority of the new codes over existing compression algorithms.
\end{abstract}

\begin{IEEEkeywords}
Data compression, fixed-width memories, table compression, Huffman coding, network switches and routers.
\end{IEEEkeywords}

\setcounter{page}{1}

\IEEEdisplaynotcompsoctitleabstractindextext

\Journal{\let\thefootnote\relax\footnotetext{
This paper solves a problem first formulated in the paper ``Compression for Fixed-Width Memories" presented at the IEEE International Symposium on Information Theory (ISIT) 2013, Istanbul, Turkey~\cite{ISITFixedWidthMemories}.
O. Rottenstreich is with Princeton University, Princeton NJ, USA (e-mail: orir@princeton.edu).
Y. Cassuto is with the Viterbi Department of Electrical Engineering, Technion -- Israel Institute of Technology, Haifa Israel (e-mail: ycassuto@ee.technion.ac.il).
\thanks{This work was supported in part by the Israel Science Foundation (ISF), and by the Intel Center for Computing Intelligence (ICRI-CI).}

}}


%
\section{Introduction}
In the best-known data-compression problem, a discrete distribution on source elements is given, and one wishes to find a representation for the source elements with minimum expected length. This problem was solved by the well-known Huffman coding~\cite{Huffman}. Huffman coding reduces the problem to a very efficient recursive algorithm on a tree, which solves it optimally. Indeed, Huffman coding has found use in numerous communication and storage applications. Minimizing the expected length of the coded sequence emitted by the source translates to optimally low transmission or storage costs in those systems. However, there is an important setup in which minimizing the expected length does {\em not} translate to optimal improvement in system performance. This setup is {\em fixed-width memory}. Information is stored in a fixed-width memory in words of $L$ bits each. A word is designated to store an entry of $d$ fields, each emitted by a source. In this case the prime objective is to fit the representations of the $d$ sources into $L$ bits, and not to minimize the expected length for each source.

The fixed-width memory setup is extremely useful in real applications. The most immediate use of it is in networking applications, wherein switches and routers rely on fast access to entries of very large tables~\cite{kirsch}. The fast access requires simple word addressing, and the large table size requires good compression. In addition to this principal application, data-centric computing applications can similarly benefit from efficient compression of large data sets in fixed-width memories.
In this paper we consider the compression of data entries with $d=2$ fields into memory words of $L$ bits, where $L$ is a parameter. This special case of two fields is motivated by switch and router tables in which two-field tables are most common. Throughout the paper we assume that the element chosen for an entry field is drawn i.i.d. from a known distribution for that field, and that elements of distinct fields are independent of each other.

Our operational model for the compression assumes that success to fit an entry to $L$ bits translates to fast data access, while failure results in a much slower access because a secondary slower memory is accessed to fit the entry. Therefore, we want to maximize the number of entries we succeed to fit, but do allow some (small) fraction of failures. Correspondingly, the performance measure we consider throughout the paper is $P_{success}$: the probability that the encoding of the two-field entry fits in $L$ bits. That is, we seek a uniquely decodable encoding function that on some entries is allowed to declare {\em encoding failure}, but does so with the lowest possible probability for the source distributions. We emphasize that we do {\em not} allow decoding errors, and all entries that succeed encoding are recovered perfectly at read time.

Maximizing $P_{success}$ is trivial when we encode the two fields jointly by a single code: we simply take the $2^L$ highest probabilities in the product distribution of the two fields, and represent each of these pairs with a codeword of length $L$. However, this solution has prohibitive cost because it requires a dictionary with size quadratic in the support of the individual distributions. For example, if each field takes one of $1000$ values, we will need a dictionary of order $10^6$ in the trivial solution. To avoid the space blow-up, we restrict the code to use dictionaries of sizes {\em linear in the distribution supports}. The way to guarantee linear dictionary sizes is by composing the entry code from two codes for the individual fields. Our task is to find two codes for the two fields, such that when the two codewords are placed together in the width-$L$ word, the entry fields can be decoded unambiguously. The design objective for the codes is to maximize $P_{success}$: the probability that the above encoding of the entry has length of at most $L$ bits. We first show by example that Huffman coding is {\em not} optimal for this design objective.
Consider a data entry with two fields, such that the value of the first field is one of $5$ possible source elements $\{a, b, c, d, e\}$, and the value of the second field is one of $3$ elements $\{x, y, z\}$. The distributions on these elements are given in the two left columns of Table~\ref{Table_Forwarding_Table_Example}.(A)-(B); the values of the two fields are drawn independently.
We encode the two fields using two codes $\sigma_1, \sigma_2$ specified in the right columns of Table~\ref{Table_Forwarding_Table_Example}.(A)-(B), respectively. The codewords of
%
$\sigma_1$, $\sigma_2$
%
%
are concatenated, and this encoding needs to be stored in the memory word. Table~\ref{Table_Forwarding_Table_Example}.(C) enumerates the $5 \cdot 3 = 15$ possible entries (combinations of values for the two fields), and for each gives its probability, its encoding, and the encoding length. The rightmost column of Table~\ref{Table_Forwarding_Table_Example}.(C) indicates whether the encoding succeeds to fit in $L=4$ bits ($\surd$), or not ($-$). The success probability of the pair of codes $(\sigma_1,\sigma_2)$ is the sum of all probabilities in rows marked with $\surd$. This amounts to
$$P_{success} =  0.20 + 0.12 + 0.08 + 0.15 + 0.09 + 0.06 +  0.08 + 0.048 + 0.032 + 0.04 + 0.03 = 0.93.$$
\SingleColMode{
\begin{table}[tbp]
\centering
\subtable[First field with code $\sigma_1$]{
\begin{tabular}{|c|c|l|}
\hline
Element   &   Prob.   &     codeword \\
\hline
$a$ & 0.4 &  00 \\
$b$ & 0.3  & 01 \\
$c$ & 0.16  & 10 \\
$d$ & 0.08  & 110 \\
$e$ & 0.06  & 111 \\
\hline
\end{tabular}
\label{Table_First_Field_Example}
}
\subtable[Second field  with code $\sigma_2$]{
\begin{tabular}{|c|c|l|}
\hline
Element   &   Prob.   &     codeword \\
\hline
$x$ & 0.5 &  0 \\
$y$ & 0.3 & 10 \\
$z$ & 0.2 & 11 \\
\hline
\end{tabular}
\label{Table_Second_Field_Example}
}
\subtable[Possible entries encoded by the encoding function $\Sigma = (\sigma_1,\sigma_2)$]{
\begin{tabular}{|p{1cm}|c|p{1.5cm}|c|c|}
\hline
Entry  & Prob. & Encoding & Width &  Width $\le (L=4)$\\
\hline
($a,x$) &  0.20 & 00 0 & 3 & $\surd$\\
($a,y$) &  0.12 & 00 10 & 4 & $\surd$\\
($a,z$) &  0.08 & 00 11 & 4 & $\surd$\\
($b,x$) &  0.15 & 01 0 & 3 & $\surd$\\
($b,y$) &  0.09 & 01 10 & 4 & $\surd$\\
($b,z$) &  0.06 & 01 11 & 4 & $\surd$\\
($c,x$) &  0.08 & 10 0 & 3 & $\surd$ \\
($c,y$) &  0.048 & 10 10 & 4 & $\surd$\\
($c,z$) &  0.032 & 10 11 & 4 & $\surd$\\
($d,x$) &  0.04 & 110 0 & 4 & $\surd$\\
($d,y$) &  0.024 & 110 10 & 5 & -\\
($d,z$) &  0.016 & 110 11 & 5 & -\\
($e,x$) &  0.03 & 111 0 & 4 & $\surd$\\
($e,y$) &  0.018 & 111 10 & 5 & -\\
($e,z$) &  0.012 & 111 11 & 5 & -\\
\hline
\end{tabular}
\label{Table_Entries_Example}
}
\caption{Example of coding for two fields. \ref{Table_First_Field_Example} and \ref{Table_Second_Field_Example} list entry distributions of the two fields, and the two codes $\sigma_1$ and $\sigma_2$ chosen for them. \ref{Table_Entries_Example} lists all combinations of values from the two fields, their encoding using the concatenation $(\sigma_1,\sigma_2)$, and the encoding lengths.  Spaces are presented for convenience, in real memory the codewords are concatenated without spaces and with trailing zeros if needed.
}
\label{Table_Forwarding_Table_Example}
\end{table}
}
It can be checked that the codes $(\sigma_1,\sigma_2)$ give better $P_{success}$ than the success probability of the respective Huffman codes, which equals 0.78.

In the sequel we design code pairs $(\sigma_1,\sigma_2)$, which we call {\em entry coding schemes}. We refer to an entry coding scheme as {\em optimal} if it maximizes $P_{success}$ among all entry coding schemes in its class. In the paper we find optimal coding schemes for two classes with different restrictions on their implementation cost: one is given in Section~\ref{sec:optimal_coding} and one in Section~\ref{Section_single_distribution}.

Finding codes that give optimal $P_{success}$ cannot be done by known compression algorithms. Also, brute-force search for an optimal code has complexity that grows exponentially with the number of source elements. Even if we consider only {\em monotone codes} -- those which assign codewords of lengths non-increasing with the element probabilities -- we can show that the number of such codes grows asymptotically as at least $\eta^n$, where $\eta=\sqrt[3]2$. An exact count of monotone codes is provided in the Appendix. The infeasibility of code search motivates the study in this paper toward devising efficient polynomial-time algorithms for finding optimal fixed-width codes.

For the case of entries with $d=2$ fields, this paper solves the problem of optimal fixed-width compression. We present an efficient algorithm that takes any pair of element distributions and an integer $L$, and outputs an entry encoder with optimal $P_{success}$. The encoder has the unique-decodability property, and requires a dictionary whose size equals the sum of the element counts for the two fields. We also solve the problem for the more restrictive setup in which {\em the same code} needs to be used for both fields. This setup is motivated by systems with a more restricted memory budget, which cannot afford storing two different dictionaries for the two fields. For both setups finding optimal codes is formulated as efficient dynamic-programming algorithms building on clever decompositions of the problems to smaller problems.

\begin{figure*}[!t]
\centering
\subfigure[Write] {
\includegraphics[trim={0.1cm 6.5cm 2 1.8cm},clip, width=0.8 \textwidth]{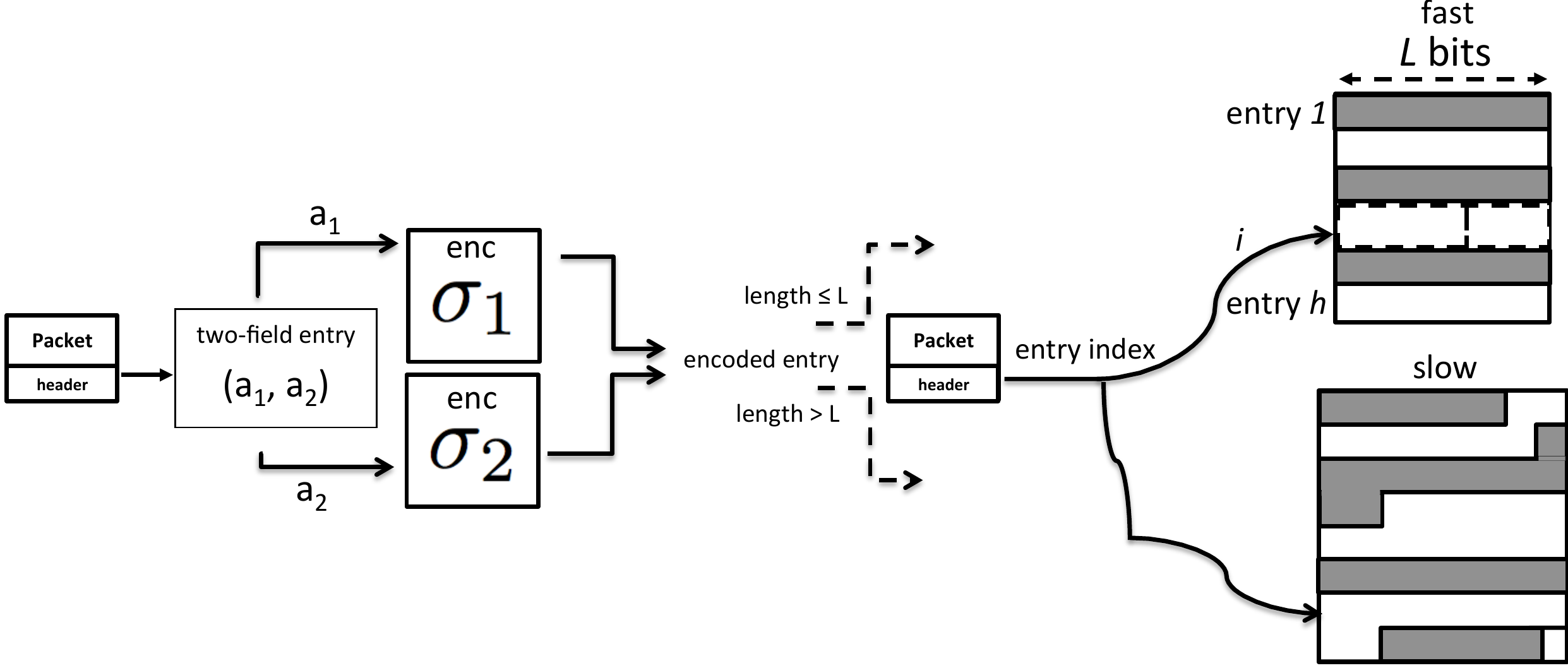}
\label{fig:implementation_write}}
\subfigure[Read] {
\includegraphics[trim={0.5cm 10.3cm 1 1.5cm},clip, width=0.8 \textwidth]{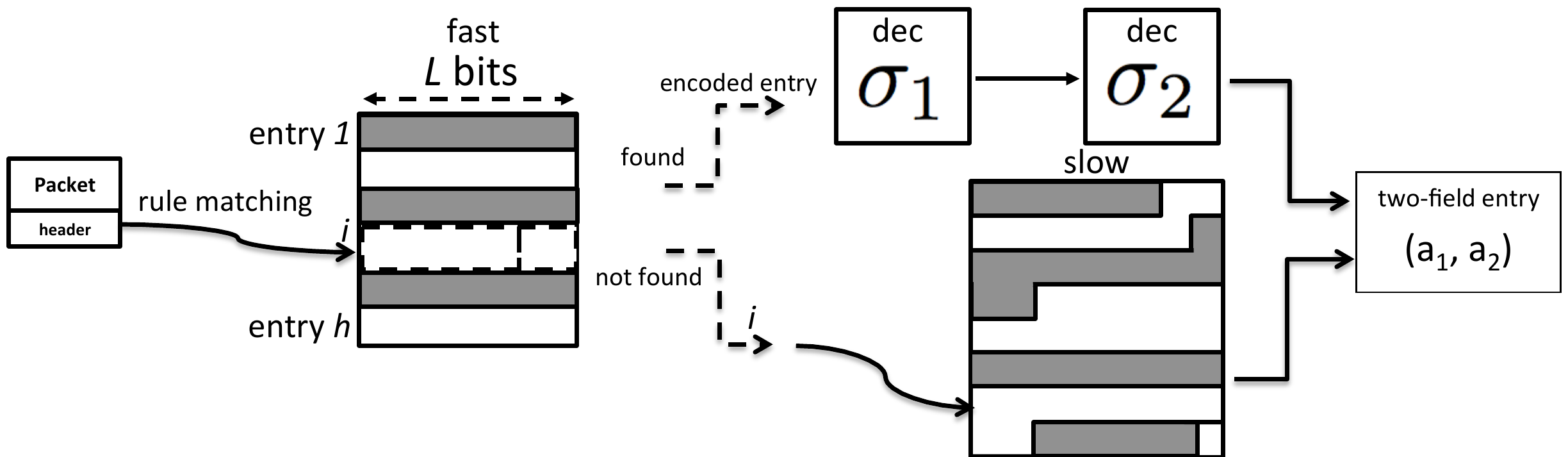}
\label{fig:implementation_read}}
\caption{
A plausible realization of a memory system with fixed-width compression. It includes a fast fixed-width (e.g., SRAM) memory with words of $L$ bits, and an additional slower (e.g. DRAM) memory. Entries that can be encoded within $L$ bits are stored in the fast memory while other entries are stored in the slow memory without the width limit. The encoding and decoding are performed based on the designed code pair $(\sigma_1, \sigma_2)$.
}
\label{fig:architecture}
\end{figure*}

Fig.~\ref{fig:architecture} shows an illustration of a plausible realization of fixed-width compression in a networking application. First, (a) illustrates the write operation where the entry is encoded and is written to the fast SRAM memory if the encoded entry has a length of at most $L$ bits. Otherwise, it is stored in the slow DRAM memory. (b) shows the read operation where encoded data is first searched in the fast SRAM memory. If it is not found, the slow DRAM memory is also accessed after the entry index is translated.

\paragraph*{Paper organization}
In Section~\ref{Section_Model}, we present the formal model for the problem of compression for fixed-width memories. Setting up the problem requires some definitions extending classical single-source compression to $2$ sources modeling the $d=2$ fields in the entry. Another important novelty of Section~\ref{Section_Model} is the notion of an encoder that is allowed to fail on some of the input pairs. In Section~\ref{sec:optimal_coding} we derive an algorithm for finding an optimal coding scheme for fixed-width compression with $d=2$ fields. There are two main ingredients to that algorithm: 1) an algorithm finding an optimal prefix code for the first field given the code for the second field, and 2) a code for the second field that is optimal for any code used in the first field. In the second ingredient it turns out that the code needs not be a prefix code, but it does need to satisfy a property which we call {\em padding invariance}. Next, in Section~\ref{Section_single_distribution} we study the special case where the two fields of an entry need to use the same code. The restriction to use one code renders the algorithm of Section~\ref{sec:optimal_coding} useless for this case. Hence we develop a completely different decomposition of the problem that can efficiently optimize a single code jointly for both fields. In Section~\ref{Section_Experimental_Results} we present empiric results showing the success probabilities of the optimal codes for different values of the memory width $L$. The results show some significant advantage of the new codes over existing Huffman codes. In addition, they compare the performance of the schemes using two codes (Section~\ref{sec:optimal_coding}) to those that use a single code (Section~\ref{Section_single_distribution}). Finally, Section~\ref{Section_conclusions} summarizes our results and discusses directions for future work.

\paragraph*{Relevant prior work}
There have been several extensions of the problem solved by Huffman coding that received attention in the literature~\cite{hu1972path,garey1974optimal,larmore1990fast,abrahams1997code,baer2006source}. These are not directly related to fixed-width memories, but they use clever algorithmic techniques that may be found useful for this problem too. The notion of fixed-width representations is captured by the work of Tunstall on variable-to-fixed coding~\cite{TUNSTALL}, but without the feature of multi-field entries. The problem of fixed-width encoding is also related to the problem of compression with low probability of buffer overflow~\cite{jelinek,Humblet81}. However, to the best of our knowledge the prior work only covers a single distribution and an asymptotic number of encoding instances (in our problem the number of encoding instances is a fixed $d$). The most directly related to the results of this paper is our prior work~\cite{JSACCompressing} considering a combinatorial version of this problem, where instead of source distributions we are given an instance with element pairs. This combinatorial version is less realistic for applications, because a new code needs to be designed after every change in the memory occupancy.

\section{Model and Problem Formulation}
\label{Section_Model}

\subsection{Definitions}
We start by providing some definitions, upon which we next cast the formal problem formulation.
Throughout the paper we assume that data placed in the memory come from known distributions, as now defined.
\begin{definition} [Element Distribution]\label{def:element_dist}
An \emph{element distribution} $(S,P)$ = $((s_{1},\dots,s_{n}),(p_{1},\dots,p_{n}))$ is characterized by an (ordered) set of elements with their corresponding positive appearance probabilities. An element of $S$ is drawn randomly and independently according to the distribution $P$, i.e.,
$\Prob{a = s_{i}} = p_{i}$, with $p_{i}> 0$ and $\Sigma_{i=1}^{n} p_{i} = 1$.
\end{definition}
Throughout the paper we assume that the elements $s_i$ are ordered in {\em non-increasing} order of their corresponding probabilities $p_i$. Given an element distribution, a {\em fixed-to-variable code} $\sigma$, which we call {\em code} for short, is a set of binary codewords $B$ (of different lengths in general), with a mapping from $S$ to $B$ called an {\em encoder}, and a mapping from $B$ to $S$ called a {\em decoder}.

In this paper we are mostly interested in coding data entries composed of {\em element pairs}, so we extend Definition~\ref{def:element_dist} to distributions of two-element data entries. For short, we refer to a data entry simply as an {\em entry}.
\begin{definition} [Entry Distribution]\label{def:entry_dist}
An \emph{entry distribution} $D= [(S_1, P_1), (S_2, P_2)]$ = $\Big[((s_{1, 1},\dots,s_{1, n_{1}}),$ $(p_{1, 1},\dots,p_{1, n_{1}}) ),$ $((s_{2, 1},\dots,s_{2, n_{2}}),$ $(p_{2, 1},\dots,p_{2, n_{2}}) )\Big]$
is a juxtaposition of two element distributions. Each field in an entry is drawn randomly and independently according to its corresponding distribution in $D$, i.e.,
$\Prob{a_1 = s_{1, i}} = p_{1, i}$ and $\Prob{a_2 = s_{2, i}} = p_{2, i}$, with  $p_{1, i}, p_{2, i}> 0$.
The numbers of possible elements in the first and second field of an entry are
$n_1 = |S_1|$ and $n_2 = |S_2|$, respectively.
\end{definition}

\begin{example}\label{exp1}
The entry distribution of the two fields illustrated in Table~\ref{Table_Forwarding_Table_Example} 
can be specified as $D =  [(S_1, P_1), (S_2, P_2)] = \Big[((a, b, c, d,e ), (0.4, 0.3, 0.16, 0.08,  0.06)) , ((x,y,z), (0.5, 0.3,0.2))\Big]$.
\end{example}
Once we defined the distributions governing the creation of entries, we can proceed to definitions addressing the coding of such entries. In all our coding schemes, entries are encoded to bit strings of fixed length $L$ (possibly with some padding), where $L$ is a parameter equal to the width of the memory in use. An entry coding scheme is defined through its {\em encoding function}.
\begin{definition} [Entry Encoding Function]\label{def:entry_encoding}
An \emph{entry encoding function} $\Sigma$ is a mapping $\Sigma: (S_1,S_2) \to \{0,1\}^{L} \cup \{\perp\}$, where $\{0,1\}^{L}$ is the set of binary vectors of length $L$, and $\perp$ is a special symbol denoting encoding failure.
\end{definition}
The input to the encoding function are two elements, one for field $1$ taken from $S_1$, and one for field $2$ taken from $S_2$. In successful encoding, the encoder outputs a binary vector of length $L$ to be stored in memory, and this vector is obtained uniquely for this input entry. The encoder is allowed to fail, in which case the output is the special symbol $\perp$. When the encoder fails, we assume that an alternative representation is found for the entry, and stored in a different memory not bound to the width-$L$ constraint. This alternative representation is outside the scope of this work, but because we know it results in higher cost (in space and/or in time), our objective here is to minimize the failure probability of the encoding function. Before we formally define the failure probability, we give a definition of the decoding function matching the encoding function of Definition~\ref{def:entry_encoding}.
\begin{definition} [Entry Decoding Function]\label{def:entry_decoding}
An \emph{entry decoding function} $\Pi$ is a mapping $\Pi:\{0,1\}^{L} \to (S_1,S_2)$, such that if $\Sigma(a_1,a_2)=b\neq\perp$, then $\Pi(b)=(a_1,a_2)$. The decoding function is limited to use $O(n_1+n_2)$ space to store the mapping.
\end{definition}
The definition of the decoding function is straightforward: it is the inverse mapping of the encoding function when encoding succeeds. The constraint of using $O(n_1+n_2)$ space for dictionaries is crucial: combining the two fields of the entry to one product distribution on $n_1\cdot n_2$ elements is both practically infeasible and theoretically uninteresting. With encoding and decoding in place, we turn to define the important measure of encoding success probability.
\begin{definition} [Encoding Success Probability]\label{def:encode_success}
Given an entry distribution $D= [(S_1, P_1), (S_2, P_2)]$ and an entry encoding function $\Sigma$, define the \emph{encoding success probability} as
\begin{equation} P_{success}(D,\Sigma) = \Pr(\Sigma(a_1,a_2)\neq \perp), \end{equation}
where the probability is calculated over all pairs of $(a_1,a_2)$ from $(S_1,S_2)$ according to their corresponding probabilities $(P_1,P_2)$.
\end{definition}
Concluding this section, we define an {\em entry coding scheme} as a pair of entry encoding function $\Sigma$ and matching entry decoding function $\Pi$, meeting the respective definitions above. Our principal objective in this paper is to find an entry coding scheme that maximizes the encoding success probability given the entry distribution and the output length $L$.

\section{Optimal Entry Coding}\label{sec:optimal_coding}
In this section we develop a general entry coding scheme that yields optimal encoding success probability under some reasonable assumptions on the entry encoding function.
\subsection{Uniquely decodable entry encoding}
The requirement from the decoding function to invert the encoding function for all successful encodings implies that the encoding function must be {\em uniquely decodable}, that is, an encoder-output bit vector $b$ represents a unique entry $(a_1,a_2)$. Constraining the decoding function to using only $O(n_1+n_2)$ space implies using two codes: $\sigma_1$ for $S_1$ and $\sigma_2$ for $S_2$, because joint encoding of $S_1,S_2$ in general requires $O(n_1\cdot n_2)$ space for decoding.
We then associate an entry encoding function $\Sigma$ with the pair of codes $(\sigma_1, \sigma_2)$.
The entry decoding function must first ``parse'' the vector $b$ to two parts: one used to decode $a_1\in S_1$ and one to decode $a_2\in S_2$.
An effective way to allow this parsing is by selecting the code  $\sigma_1$ for $S_1$ to be a {\em prefix code}, defined next.
\begin{definition} [Prefix Code]
\label{Definition_Prefix_Code}
For a set of elements $S$, a code $\sigma$ is called a \emph{prefix code} if in its codeword set $B$ no codeword is a prefix (start) of any other codeword.
\end{definition}
Thanks to using a prefix code for $S_1$, the decoder can read the length-$L$ vector $b$ from left to right and unequivocally find the last bit representing $a_1$, and in turn the remaining bits representing $a_2$. Hence in the remainder of the section we restrict the entry encoding functions to have a constituent prefix encoder for their first field. This restriction of the encoder to be prefix is reminiscent of the classical result in information theory by Kraft-McMillan~\cite{AbramsonIT} that prefix codes are sufficient for optimal fixed-to-variable uniquely decodable compression. In our case we cannot prove that encoding functions with a prefix code for $S_1$ are always sufficient for optimality, but we also could not find better than prefix encoding functions that attain the $O(n_1+n_2)$ space constraint at the decoder. To be formally correct, we call an entry coding scheme optimal in this section if it has the maximum success probability among coding schemes whose first code is prefix.


For unique decodability in the second field also, we use a special code property we call {\em padding invariance}.
\begin{definition} [\PInv Code]
A code is called a {\pinv code} if  after eliminating the trailing (last) zero bits from its codewords $B$  (possibly including the empty codeword $\emptyset$ of length 0), $|B|$ distinct binary vectors are obtained.
\end{definition}

\begin{example}
Consider the following codes: $\sigma$ with a set of codewords $\{0, 11, 100, 101\}$,  $\sigma'$ with $\{\emptyset, 1, 110\}$ and  $\sigma''$ with $\{1, 10, 00\}$.
The code $\sigma$ is a prefix code since none of its codewords is a prefix of another codeword. It is also a \pinv code since after eliminating the trailing zeros, the four distinct codewords $\emptyset$ (the empty codeword), $11,1$ and $101$ are obtained.
While $\sigma'$ is also a  \pinv code (the codewords $\emptyset, 1, 11$ are distinct), it is not a prefix code since 1 is a prefix of 110 and $\emptyset$ is a prefix of 1 and 110.
The code $\sigma''$  is neither a prefix code (1 is a prefix of 10) nor a \pinv code (the codeword 1 is obtained after eliminating the trailing zero bits of 1 and 10).
\end{example}

We now show how a prefix code $\sigma_1$ and a padding-invariant code $\sigma_2$ are combined to obtain a uniquely-decodable entry encoding function. The following is not yet an explicit construction, but rather a general construction method we later specialize to obtain an optimal construction.
\setcounter{construction}{-1}
\begin{construction}\label{cnst:prefix_pinv}
Let $\sigma_1$ be a prefix code and $\sigma_2$ be a padding-invariant code. The vector $b=\Sigma(a_1,a_2)$ output by the entry encoding function $\Sigma = (\sigma_1,\sigma_2)$ is defined as follows. The entry encoding function first maps $a_1$ to a word of $\sigma_1$ and $a_2$ to a word of $\sigma_2$. Then the vector $b$ is formed by concatenating the codeword of $\sigma_2$ to the right of $\sigma_1$'s codeword, and padding the resulting binary vector with zeros on the right to get $L$ bits total. If the two codewords exceed $L$ bits before padding, the entry encoding function returns $\perp$.
\end{construction}
We show that because $\sigma_1$ is a prefix code while $\sigma_2$ is a \pinv code (not necessarily prefix), the entry encoding function of Construction~\ref{cnst:prefix_pinv} is uniquely decodable.
It is easy to see also that padding invariance of the second code is necessary to guarantee that.
\begin{property}\label{property_decodability_general_case}
Let  $\Sigma = (\sigma_1,\sigma_2)$ be an entry encoding function specified by Construction~\ref{cnst:prefix_pinv}.
Let $x=(x_1,x_2), y=(y_1,y_2)$ be two entries
composed of elements $x_1, y_1 \in S_1$, $x_2, y_2 \in S_2$.
For $\cdot$ representing concatenation, let $\Sigma(x) = \sigma_1(x_1)  \cdot \sigma_2(x_2) \cdot 0 \cdot \ldots \cdot 0$ and
$\Sigma(y) = \sigma_1(y_1) \cdot \sigma_2(y_2)  \cdot 0 \cdot \ldots \cdot 0$ be the two length-$L$ vectors output by the entry encoding function $\Sigma$.
If $\Sigma(x) = \Sigma(y)$, then necessarily $x=y$, i.e., $x_1 = y_1, x_2 = y_2$.
\end{property}
To show the correctness of this property we explain why the decoding of a coded entry is unique.

\bp
We first explain that there is a single element $u_1 \in S_1$ whose encoding $\sigma_1(u_1)$ is a prefix of $\Sigma(x) = \Sigma(y)$. Assume the contrary, and let $u, u' \in S_1$ be two (distinct) elements such that $\sigma_1(u)$ and $\sigma_1(u')$  are both prefixes of $\Sigma(x) = \Sigma(y)$. It then follows that either $\sigma_1(u)$ is a prefix of $\sigma_1(u')$ or $\sigma_1(u')$ is a prefix of $\sigma_1(u)$. Since $u \ne u'$ and $\sigma_1$ is a prefix code, both options are impossible, and we must have that there is indeed a single element $u_1 \in S_1$ whose encoding $\sigma_1(u_1)$ is a prefix of $\Sigma(x) = \Sigma(y)$.
Thus necessarily $\sigma_1(x_1) = \sigma_1(y_1)$ and $x_1 = y_1$. By
eliminating the first identical bits that stand for $\sigma_1(x_1) = \sigma_1(y_1)$ from $\Sigma(x) = \Sigma(y)$, we are left with the encoding of the second field, possibly including zero trailing bits.
By the properties of the \pinv code $\sigma_2$, it has at most a single element that maps to the bits remaining after eliminating any number of the trailing zero bits.
We identify this codeword as $\sigma_2(x_2) = \sigma_2(y_2)$, and deduce the only possible element $x_2 = y_2$ of this field that has this codeword.
\ep

To decode an encoded entry composed of the encodings $\sigma_1(x_1)$, $\sigma_2(x_2)$, possibly with some padded zero bits, a decoder can simply identify the single element $(u_1 = x_1) \in S_1$ whose encoding $\sigma_1(u_1=x_1)$ is a prefix of the encoded entry.
As $\sigma_1$ is a prefix code, there cannot be more than one such element.
The decoder can then identify the beginning of the second part of the encoded entry, which encodes a unique codeword of $\sigma_2$ with some trailing zero bits.
This is necessarily $\sigma_2(x_2)$ from which $x_2$ can be derived.

Toward finding an entry coding scheme with maximal encoding-success probability, we would now like to find codes $\sigma_1$ and $\sigma_2$ for Construction~\ref{cnst:prefix_pinv} that are optimal given an entry distribution $D$ and the output length $L$. Since the direct joint computation of optimal $\sigma_1$ and $\sigma_2$ codes is difficult, we take a more indirect approach to achieve optimality. 1) we first derive an efficient algorithm to find an optimal prefix code for $S_1$ \underline{given a code for $S_2$}, and then 2) we find a (padding-invariant) code for $S_2$ that is universally optimal for any code used for $S_1$. This way we reduce the joint optimization of $\sigma_1$,$\sigma_2$ (hard) to a conditional optimization of $\sigma_1$ given $\sigma_2$ (easier). We prove that this conditionally optimal $\sigma_1$ is also unconditionally optimal. To do so, we show that the code $\sigma_2$ assumed in the conditioning is optimal for any choice of $\sigma_1$. These two steps are detailed in the next two sub-sections, and later used together to establish an optimal coding scheme.

\begin{algorithm}[b!]
	\label{algorithm_conditional_code}
	\SetInd{.03in}{.03in}
	\SetKwInOut{Input}{input}
	\SetKwInOut{Output}{output}
	\caption{Optimal Conditional Prefix Code}
	\Input{Entry distribution $D$, memory width $L$, code $\sigma_2$}
	\Output{Conditionally optimal prefix code $\sigma_1$}
	\underline{{\em initialization}:}\\
	\lForEach{$N \ge 0$}{
		$F(k=0,N) = 0, Q(N,k) = \big(\big)$
	}
	\lForEach{$N < 0$}{
		$F(k = 0,N) = - \infty$
	}
	\underline{{\em intermediate solutions}:}\\
	\For{$k=1:n_1$}{
		\For{$N=0:2^L$}{
			$F(k,N) =   \max \bigg( \max_{\ell_0 \in [1, L]} \, \Big( F(k-1, N - N_{\ell_0}) + p_{1, k} \cdot  \sum _{i=1}^{n_2}  p_{2, i} \cdot I \Big[ \ell_0 + \ell (\sigma_2(s_{2,i}) )   \le L \Big] \Big) ~,~ F(k-1,N) \bigg)$\\
		\eIf{\emph{outer max attained by first argument}}{
	$Q(k,N)$ =  $(Q( k-1, N - N_{\ell_0^{*}}), \ell_0^{*})$, where $\ell_0^{*}$ is the $\ell_0$ that attains the inner max}{
$Q(k,N)=(Q(k-1,N), \infty)$
}

		}
	}
	Calculate prefix code $\sigma_1$ based on codeword lengths given by $Q(k = n_1, N = 2^{L})$\\
	\underline{{\em output results}:}\\
	
	Encoding success probability $P_{success} = F(k = n_1, N = 2^{L})$\\
	Optimal conditional prefix code $\sigma_1$
\end{algorithm}

\subsection{An optimal prefix code for $S_1$  conditioned on a code for $S_2$}\label{subsec:cond_opt}

Our objective in this sub-section is to find a prefix code $\sigma_1$ that maximizes the encoding success probability given a code $\sigma_2$ for the second field. We show constructively that this task can be achieved with an efficient algorithm.

We denote $W_i = \left \lceil \log_2(n_i) \right \rceil$ for $i \in \{1,2\}$ and $W = W_1 + W_2$. 
Finding an optimal coding scheme when $L \ge W$ is easy.
In that case we could allocate $W_1$ bits to $\sigma_1$ and $W_2$ bits to $\sigma_2$ and apply two independent fixed-length codes with a total of $W$ bits and obtain a success probability $1$. Thus we focus on the interesting case where $L \le W-1$.

Given a code $\sigma_2 = \sigma$ for $S_2$, we show a polynomial-time algorithm that finds an optimal conditional prefix code $\sigma_1$ for $S_1$. This code $\sigma_1$ will have an encoding function maximizing the probability
$P_{success}(D, \Sigma = (\sigma_1,\sigma_2= \sigma))$ given $\sigma_2 = \sigma$, when $\sigma_1$  is restricted to be prefix. Then, we also say that
$\Sigma = (\sigma_1,\sigma_2)$ with its corresponding decoding function $\Pi$ is an optimal conditional coding scheme.

To build the code $\sigma_1$ we assign codewords to $n'$ elements of $S_1$, where $n'\leq n_1$. Clearly, all such codewords have lengths of at most $L$ bits.
For a (single) binary string $x$, let $\ell(x)$ denote the {\em length} in bits of $x$.
Since for every element $a \in S_1$ which is assigned a codeword the code $\sigma_1$ satisfies $\ell (\sigma_1(a)) \le L$, it holds that $2^{- \ell (\sigma_1(a))}$ must be an integer multiple of $2^{-L}$.
We define the {\em weight} of a codeword of length $\ell_0$ as the number of units of $2^{-L}$ in $2^{- \ell_0}$, denoted by
$N_{\ell_0} = 2^{- \ell_0} / 2^{-L} = 2^{L - \ell_0}$. The $n_1-n'$ elements of $S_1$ not represented by $\sigma_1$ are said to have length $\ell_0=\infty$ and codeword weight zero.
A prefix code exists with prescribed codeword lengths if they satisfy Kraft's inequality~\cite{BooksDaglib0016881}. In our terminology, this means that the sum of weights of the codewords of $\sigma_1$ need be at most $2^{L}$.
\begin{definition}\label{def:F_k_N}
Consider entries composed of an element from the first (highest probability) $k$ elements of $S_1$ (for $k \in [0,n_1]$), and an arbitrary element of $S_2$.
For $N \in [0, 2^{L}]$ and $k \in [0,n_1]$, we denote by $F(k,N)$ the maximal sum of probabilities of such entries that can be encoded successfully by a prefix code $\sigma_1$ whose sum of weights for the first $k$ codewords is at most $N$. Formally,
\SingleColMode{
\begin{align}
F(k,N) =   &\max_{\sigma_1 : \parent{\sum _{j=1}^{k} {N_{\ell (\sigma_1(s_{1,j}))}}  \le N} } \, \Bigg( \sum _{i=1}^{k} \sum _{j=1}^{n_2} p_{1, i} \cdot p_{2, j}
 \cdot \emph{I} \bigg[ \ell (\sigma_1(s_{1,i}) ) + \ell (\sigma_2(s_{2,j}))\le L \bigg] \Bigg), \label{eq:F_opt}
\end{align}
}
where $\emph{I}[\cdot]$ is the indicator function. Note that $F(k,N)$ depends on the conditioned $\sigma_2$, but we keep this dependence implicit to simplify notation.
\end{definition}
The following theorem relates the maximal success probability of a conditional coding scheme and the function $F(k,N)$.
\begin{theorem}\label{theorem_function_F_k_N_optimal_success_probability}
The maximal success probability of a conditional coding scheme is given by
\begin{align}
\max_{\sigma_1} \, P_{success}(D, \Sigma = (\sigma_1,\sigma_2=\sigma_2))
= F(k = n_1, N = 2^{L}). \nonumber
\end{align}
\end{theorem}
\bp
To satisfy  Kraft's inequality we should limit the sum of weights $N$ to $2^{L}$. In addition,  the success probability of the coding scheme with an encoding function $\Sigma$
is calculated based on entries with any number of the $n_1$ elements of $S_1$.
\ep

We next show how to compute $F(k,N)$ efficiently for all $k,N$, in particular for $k = n_1$, $N = 2^{L}$ that yield the optimal conditional $\sigma_1$. To do that, we use the following recursive formula for $F(k,N)$.
First note the boundary cases $F(k = 0,N) = 0$ for $N \ge 0$ and $F(k = 0,N) = - \infty$ for $N < 0$ (this means an invalid code).
We can now present the formula of $F(k, N)$
that calculates its values for $k$ based on the values of the function for $k-1$ and $N'\leq N$.

\TwoColMode{
\begin{lemma}\label{lemma_function_F_k_N_recursive_formula}
The function $F(k,N)$ satisfies for $N \ge 0, k \ge 1$
\begin{align}
F(k,N) =   &\max \Bigg( \max_{\ell_0 \in [1, L]} \, \bigg( F(k-1, N - N_{\ell_0})
+ p_{1, k} \cdot  \sum _{i=1}^{n_2}  p_{2, i} \cdot I \Big[ \ell_0 + \ell (\sigma_2(s_{2,i}) )   \le L \Big] \bigg), F(k-1,N) \Bigg).\end{align}
\end{lemma}
}
\SingleColMode{
\begin{lemma}\label{lemma_function_F_k_N_recursive_formula}
The function $F(k,N)$ satisfies for $N \ge 0, k \ge 1$
\begin{align}
F(k,N) =   &\max \Bigg( \max_{\ell_0 \in [1, L]} \, \bigg( F(k-1, N - N_{\ell_0})
+ p_{1, k} \cdot  \sum _{i=1}^{n_2}  p_{2, i} \cdot I \Big[ \ell_0 + \ell (\sigma_2(s_{2,i}) )   \le L \Big] \bigg), F(k-1,N) \Bigg).\label{eq:recur_F}\\
\end{align}
\end{lemma}
}
\bp
The optimal code that attains $F(k,N)$ either assigns a codeword to $s_{1,k}$ or does not. The two arguments of the outer max function in~\eqref{lemma_function_F_k_N_recursive_formula} are the respective  success probabilities for these two choices. In the former case we consider all possible lengths of the codeword of $s_{1,k}$. A codeword length of $\ell_0$ reduces the available sum of weights for the first $k-1$ elements by $N_{\ell_0} = 2^{L - \ell_0}$. In addition, an entry $(s_{1,k}, s_{2,i})$
contributes to the success probability the value $p_{1, k} \cdot p_{2, i}$ if its encoding width (given $\ell_0$) is at most $L$. In the latter case the $k^\text{th}$ element does not contribute to the success probability and has no weight, hence $F(k,N)=F(k-1,N)$ in this case.
\ep

Finally, the pseudocode of the dynamic-programming algorithm that finds the optimal conditional code based on the above recursive formula is given in Algorithm~\ref{algorithm_conditional_code}.
It  iteratively calculates the values of $F(k,N)$. It also uses a vector $Q(k,N)$ to represent the codeword lengths for the first at most $k$ elements of $S_1$ in a solution achieving $F(k,N)$.

\emph{Time Complexity}: 
By the above description, there are $n_1$ iterations and in each $O(2^{L})$ values are calculated, each by considering $O(L)$ sums of $n_2$ elements. It follows that the time complexity of the algorithm is $O(n_1 \cdot 2^{L}\cdot L\cdot n_2)= O(n_1^2 \cdot n_2^2 \cdot L)$, which is polynomial in the size of the input. The last equality follows from the fact that $L<\lceil \log_2(n_1)\rceil + \lceil \log_2(n_2)\rceil$ in a non-trivial instance where some entries fail encoding.

\subsection{A universally optimal code for $S_2$}
We now develop the second component required to complete an unconditionally optimal entry coding scheme: a code for the $S_2$ elements that is optimal for any code $\sigma_1$ used for the $S_1$ elements.

To that end we now define the natural notion of a monotone coding scheme, in which higher-probability elements are assigned codeword lengths shorter or equal to lower-probability ones.
\begin{definition} [Monotone Coding Scheme]
	A coding scheme with an encoding function $\Sigma$ of an entry distribution $D= [(S_1, P_1), (S_2, P_2)]$ is called
	\emph{monotone} if is satisfies
	
	\noindent \emph{(i)} for $j \in \{1,2\}$, $i < i'$ implies that $\ell (\sigma_j(s_{j,i}) ) \le \ell (\sigma_j(s_{j,i'}))$ for two elements $s_{j,i}, s_{j,i'} \in S_j$.
	
	\noindent \emph{(ii)} for $j \in \{1,2\}$, the elements that are not assigned a codeword in $\sigma_j$ are the last elements in $S_j$ (if any).
\end{definition}
It is an intuitive fact that without loss of generality an optimal entry coding scheme can be assumed monotone. We prove this in the following.
\begin{lemma}\label{lemma_codewords_lengths_order}
	For any entry distribution  $D= [(S_1, P_1), (S_2, P_2)]$, and any memory width $L \ge 1$, there exists
	a \emph{monotone optimal} coding scheme.
\end{lemma}
\bp 
We show how to build an optimal monotone coding scheme based on any optimal coding scheme with an encoding function $\Sigma$.
For $j \in \{1,2\}$ consider two arbitrary indices $i, i'$ that satisfy $i < i'$. Then necessarily $p_{j, i} \ge p_{j, i'}$.
If codewords are assigned to the two elements and $\ell (\sigma_j(s_{j,i}) ) > \ell (\sigma_j(s_{j,i'}))$, we can replace $\sigma_j$ by a new code
obtained by permuting the two codewords of $s_{j,i}, s_{j,i'}$.
With this change, an entry $(a_1, a_2)$ with $a_j = s_{j,i}$ is encoded successfully after the change if the corresponding  entry with $a_j = s_{j,i'}$ was encoded successfully before the change.
Then, we deduce that such a change cannot decrease $P_{success}$, and the result follows. In addition, if some elements are not assigned codewords, the same argument shows that these elements should be the ones with the smallest probabilities.
\ep
The optimality of the code to be specified for $S_2$ will be established by showing it attains the following upper bound on success probability.
\begin{prop}\label{prop:Psuc_ub}
	Given any code $\sigma_1$, the encoding success probability of any entry encoding function is bounded from above as follows.
\begin{equation}\label{eq:Psuc_ub}
P_{success} \leq \sum _{i=1}^{n'}  p_{1, i} \sum _{j=1}^{2^{L-\ell (\sigma_1(s_{1,i}) )}}  p_{2, j},
\end{equation}
where $n'$ is the highest index of an element $s_{1,i}$ that is assigned a codeword by $\sigma_1$.
\end{prop}
\bp
First by the monotonicity property proved in Lemma~\ref{lemma_codewords_lengths_order}, $\sigma_1$ assigns codewords to elements $s_{1,i}$ with indices $i\in\{1,\ldots,n'\}$, for some $n'\leq n_1$. Hence for indices greater than $n'$ the success probability is identically zero. Given an element $s_{1,i}$ with a $\sigma_1$ codeword of length $\ell_i$, at most $2^{L-\ell_i}$ elements of $S_2$ can be successfully encoded with it in an entry. So the inner sum in~\eqref{eq:Psuc_ub} is the maximal success probability given $s_{1,i}$. Summing over all $i$ and multiplying by $p_{1,i}$ gives the upper bound.
\ep
It turns out that there exists a padding-invariant code $\hat{\sigma}_2$ that attains the upper bound of Proposition~\ref{prop:Psuc_ub} with equality for any code $\sigma_1$. For the ordered set $S_2=(s_{2,1},\ldots,s_{2,n_2})$, let the encoder of $\hat{\sigma}_2$ map $s_{2,j}$ to the shortest binary representation of $j-1$ for $j\geq 2$, and to $\emptyset$ for $j=1$. The binary representation is put in the codeword from left to right, least-significant bit (LSB) first. Then we have the following.
\begin{prop}\label{prop:Psuc_att}
	Given any code $\sigma_1$, the encoding success probability of $\hat{\sigma}_2$ is
\begin{equation}\label{eq:Psuc_attain}
P_{success} = \sum _{i=1}^{n'}  p_{1, i} \sum _{j=1}^{2^{L-\ell (\sigma_1(s_{1,i}) )}}  p_{2, j},
\end{equation}
where $n'$ is the highest index of an element $s_{1,i}$ that is assigned a codeword by $\sigma_1$.
\end{prop}
\bp
If the codeword of $\sigma_1$ uses $\ell$ bits, there are $L-\ell$ bits left vacant for $\sigma_2$. The mapping specified for $\hat{\sigma}_2$ allows encoding successfully the first $2^{L-\ell}$ elements of $S_2$, which gives the stated success probability.
\ep
In particular, when the encoding of $\sigma_1$ has length $L$, the single element $s_{2,1}$ is encoded successfully by the empty codeword $\emptyset$. Other examples are the two codewords ($\emptyset$, 1) when $\ell=L-1$, and the four codewords ($\emptyset$, 1, 01, 11) when $\ell=L-2$. It is clear that the code $\hat{\sigma}_2$ is padding invariant, because its codewords are minimal-length binary representations of integers. Now we are ready to specify the optimal entry coding scheme in the following.
\begin{construction}\label{cnst:opt_prefix_pinv}
Given $L$ and $D$, let $\hat{\sigma}_1$ be the prefix code obtained by applying Algorithm~\ref{algorithm_conditional_code} on the code $\hat{\sigma}_2$. Then the entry encoding function $\hat{\Sigma}=(\hat{\sigma}_1,\hat{\sigma}_2)$ is defined by applying Construction~\ref{cnst:prefix_pinv} on $\hat{\sigma}_1$,$\hat{\sigma}_2$.
\end{construction}

\begin{algorithm}[t!]
	\label{algorithm_optimal_encoding_function}
	\SetInd{.03in}{.03in}
	\SetKwInOut{Input}{input}
	\SetKwInOut{Output}{output}
	\caption{}
	\Input{Entry distribution $D$, memory width $L$}
	\Output{Optimal coding scheme}
	\underline{{\em calculation}:}\\
	Calculate padding invariant code $\sigma_2 = \sigma^0_2$ by assigning the $n_2$ elements the codewords that correspond to $[0,n_2-1]$\\
	Calculate prefix code $\sigma_1$ by Algorithm~\ref{algorithm_conditional_code} given the code $\sigma_2$\\
	\underline{{\em output results}:}\\
	An optimal coding scheme with an encoding function $\Sigma = (\sigma_1, \sigma_2)$
\end{algorithm}

\begin{theorem}\label{theorem_algorithm_optimality}
For any entry distribution $D$ and a memory width $L$, Construction~\ref{cnst:opt_prefix_pinv} gives an optimal entry coding scheme, that is, a coding scheme that maximizes the success probability among all uniquely-decodable coding schemes with a prefix code in the first field.
\end{theorem}
From Theorem~\ref{theorem_algorithm_optimality} we can readily obtain an efficient algorithm finding an optimal two-field entry encoding, which is given in Algorithm~\ref{algorithm_optimal_encoding_function}. Optimality is proved up to the assumption that the first code is a prefix code. It is not clear how one can obtain better codes than those with a prefix $\sigma_1$, while keeping the dictionary size $O(n_1+n_2)$.

For the special case when $\sigma_2$ is $\hat{\sigma}_2$, the recursive formula for the calculation of the function $F(k,N)$ can be simplified as follows.
\begin{lemma}\label{lemma_function_F_k_N_recursive_simple_formula}
When $\sigma_2$ is $\hat{\sigma}_2$, the function $F(k,N)$ satisfies for $N \ge 0, k \ge 1$
\begin{align}
F(k,N) =
 \max \Bigg( \max_{\ell_0 \in [1, L]} \, \bigg( F(k-1, N - N_{\ell_0})
+ p_{1, k} \cdot  \sum _{i=1}^{min(n_2,  2^{L - \ell_0} )}  p_{2, i}  \bigg), F(k-1,N) \Bigg).\label{eq:recur_F_simple}
\end{align}
\end{lemma}
It is easily seen that~\eqref{eq:recur_F_simple} is obtained from~\eqref{eq:recur_F} by replacing the indicator function with the partial sum that accommodates all the $S_2$ elements that have a short enough representation to fit alongside the $S_1$ element.


The following example illustrates Construction~\ref{cnst:opt_prefix_pinv} on the entry distribution from the Introduction.
\begin{example}
\label{example_detailed}
Consider the entry distribution
$D = \Big[((a, b, c, d,e ), (0.4, 0.3, 0.16, 0.08,  0.06)) , ((x,y,z), (0.5, 0.3,0.2))\Big]$
from Table~\ref{Table_Forwarding_Table_Example} with $n_1=5, n_2=3$. The width parameter is $L=4$. For the ordered set $S_2=(x,y,z)$, we select the code $\hat{\sigma}_2$ by mapping
$s_{2,1}$ to $\emptyset$  and $s_{2,j}$ (for $j=2,3$) to the shortest binary representation of $j-1$. Then,
$\hat{\sigma}_2(x) = \emptyset, \hat{\sigma}_2(y) = 1, \hat{\sigma}_2(z) = 01$ and
$\ell(\hat{\sigma}_2(x)) = 0, \ell(\hat{\sigma}_2(y)) = 1, \ell(\hat{\sigma}_2(z)) = 2$.
The code $\hat{\sigma}_2(x)$ is a padding-invariant code.
To get
the prefix code
 $\hat{\sigma}_1$ we apply Algorithm~\ref{algorithm_conditional_code} on the code $\hat{\sigma}_2$.
We recursively calculate the values of $F(k,N)$ and $Q(k,N)$ for $k \in [1,n_1 = 5]$, $N \in [0,2^L=16]$. In particular, for each value of $k$ the values are calculated based on the previous value of $k$. The values are listed in Table~\ref{Table_detailed_example}. Each column describes a different value of $k$.
(Whenever values of $F$ and $Q$ are not shown, a specific value of $N$ does not improve the probability achieved for a smaller value of $N$ in the same column.)
The value of $N$ implies a restriction on the values of the codeword lengths. If the lengths of the codewords are described by a set $Q$, they must satisfy $\Sigma_{\ell \in Q} 2^{-\ell} / 2^{-L} \le N$, i.e. $\Sigma_{\ell \in Q} 2^{-\ell} \le N / 16$.

We first explain the values for $k=1$, considering the contribution to the success probability of data entries with $a \in S_1$  as the first element. This happens w.p.  $p_{1,1} = 0.4$.
For $N=1$ we must have $Q(1,1) = (4)$, i.e. the element a is assigned a codeword of length 4. Then, there is a single pair $(a,x)$ that can be encoded successfully
and $P_{success}$ is given by $F(1,1) $ is $0.4 \cdot 0.5 = 0.2$. Likewise, for $N=2$, we can have a codeword of length 3 for $a$ and the two pairs $(a,x), (a,y)$ can be encoded within $L=4$ bits, such that $F(1,2)  = 0.4 \cdot (0.5 + 0.3) = 0.32$. If $N=3$ we cannot further decrease the codeword length and improve the success probability. For $N=4$ we can have a codeword length of 2 bits, as described by $Q(1,4) = (2)$. This enables encoding successfully the three pairs $(a,x), (a,y) , (a,z) $   with a success probability of 0.4 as given by $F(1,4)$.
The values for larger values of $k$ are calculated in a similar manner based on the recursive formulas.
The optimal  codeword lengths for  $\hat{\sigma}_1$ are given  by $Q(k = n_1 = 5,N = 2^L = 16) =(2,2,2,3,3)$.
This enables to encode successfully all pairs besides $(d,z), (e,z)$, achieving
 $P_{success} =  0.972$ as given by $F(k = 5,N = 16)$.

Finally, by applying Construction~\ref{cnst:prefix_pinv} on $\hat{\sigma}_1$,$\hat{\sigma}_2$ we obtain the
entry encoding function $\hat{\Sigma}=(\hat{\sigma}_1,\hat{\sigma}_2)$.
\end{example}

    \begin{table}[t!]
        \centering
        \caption{The values of $F(k,N)$ (top of each pair in the table) and $Q(k,N)$ (bottom) for $k \in [1,n_1 = 5]$, $N \in [0,2^L=16]$ for Example~\ref{example_detailed} with $L=4, n_1 = 5, n_2 = 3$. The optimal value of $P_{success}$ is given by $F(5,16)$ and the codeword lengths for $\sigma_1$ by $Q(5,16)$.}
        \begin{tabular}{|c| c c c c c |}
            \hline
                        & $k=1$   & $k=2$  &  $k=3$        & $k=4$         &  $k=5$  \\
            \hline
            $N=0$ &   0         &    0           &  0              & 0   & 0\\
                       &   (-)      &     (-,-)       &   (-,-,-)     &  (-,-,-,-)               & (-,-,-,-,-)  \\
            $N=1$ &   0.2      &     0.2       &  0.2           &  0.2               &  0.2 \\
                        &   (4)      &     (4,-)       &   (4,-,-)     &  (4,-,-,-)               & (4,-,-,-,-)  \\
            $N=2$ &   0.32    &     0.35       &   0.35      &   0.35              & 0.35 \\
                        &   (3)      &     (4,4)       &   (4,4,-)   &   (4,4,-,-)             & (4,4,-,-,-) \\
            $N=3$ &              &     0.47       &   0.47       &    0.47             &  0.47 \\
                        &              &     (3,4)       &   (3,4,-)   &    (3,4,-,-)              &  (3,4,-,-,-)\\
            $N=4$ &   0.4      &     0.56       &  0.56        &    0.56             & 0.56\\
                        &   (2)      &     (3,3)       &   (3,3,-)   &     (3,3,-,-)             & (3,3,-,-,-)\\
            $N=5$ &              &                   &    0.64      &       0.64         &  0.64\\
                        &              &                   &    (3,3,4)   &     (3,3,4,-)           & (3,3,4,-,-)\\
            $N=6$ &              &   0.64         &    0.688     &   0.688            &  0.688\\
                        &              &     (2,3)       &    (3,3,3)   &    (3,3,3,-)          & (3,3,3,-,-)\\
            $N=7$ &              &                    &      0.72     &    0.728         &  0.728\\
                        &              &                    &      (2,3,4)  &    (3,3,3,4)        & (3,3,3,4,-)\\
            $N=8$ &              &     0.7          &    0.768       &     0.768        & 0.768\\
                        &              &     (2,2)       &     (2,3,3)      &   (2,3,3,-)        & (2,3,3,-,-)\\
            $N=9$ &              &                    &    0.78            &   0.808        &  0.808\\
                        &              &                    &      (2,2,4)          &   (2,3,3,4)     & (2,3,3,4,-)\\
            $N=10$ &            &                    &    0.828           &  0.832         & 0.838\\
                         &              &                  &      (2,2,3)          &   (2,3,3,3)      & (2,3,3,4,4)\\
            $N=11$ &            &                    &                      &      0.868             & 0.868 \\
                         &              &                   &               &        (2,2,3,4)          &  (2,2,3,4,-)\\
            $N=12$ &            &                    &    0.86           &     0.892       &  0.898\\
                         &              &                  &    (2,2,2)          &    (2,2,3,3)      & (2,2,3,4,4)\\
            $N=13$ &            &                    &            &          0.9           &  0.922\\
                         &              &                   &            &       (2,2,2,4)              & (2,2,3,3,4)\\
            $N=14$ &            &                    &            &        0.924             & 0.94\\
                         &              &                  &              &      (2,2,2,3)             & (2,2,3,3,3)\\
            $N=15$ &            &                    &             &                    & 0.954\\
                         &              &                   &            &                      & (2,2,2,3,4)\\
            $N=16$ &            &                    &           &        0.94               & 0.972\\
                         &              &                   &           &        (2,2,2,2)              &  (2,2,2,3,3)\\
            \hline
            \end{tabular}
        \label{Table_detailed_example}
    \end{table}


\section{Optimal Entry Encoding with the Same Code for Both Fields}
\label{Section_single_distribution}
In this section we move to study the problem of entry coding schemes for the special case where we require that both fields use the same code. It is commonly the case that both fields have the same element distribution, and then using one code instead of two can cut the dictionary storage by half. In practice this can offer a significant cost saving. Throughout this section we thus assume that the fields have the same distribution, but the results can be readily extended to the case where the distributions are different and we still want a single code with optimal success probability. Formally, in this section our problem is to efficiently design a single code $\sigma$ that offers optimal encoding success probability in a width-$L$ memory.

In the special case of a single distribution we have an element distribution $(S,P)$ = $((s_{1},\dots,s_{n}),(p_{1},\dots,p_{n}))$, and the entry distribution is $D= [(S, P), (S, P)]$. Now the space constraint for the decoder (to hold the dictionary) is to be of size at most $n$. This means that we need to find one code $\sigma$ for $(S,P)$ that will be used in both fields. To be able to parse the two fields of the entry, $\sigma$ needs to be a prefix code.

\subsection{Observations on the problem}\label{subsec:single_observ}
Before moving to solve the problem, it will be instructive to first understand the root difficulty in restricting both fields to use the same code. If we try to extend the dynamic-programming solution of Section~\ref{sec:optimal_coding} to the  single-distribution,single-code case, we get the following maximization problem
\SingleColMode{
\begin{align}
F(k,N) =   &\max_{\sigma : \parent{\sum _{j=1}^{k} {N_{\ell (\sigma(s_{j}))}}  \le N} } \, \Bigg( \sum _{i=1}^{k} \sum _{j=1}^{n} p_{i} \cdot p_{j}
 \cdot \emph{I} \bigg[ \ell (\sigma(s_{i}) ) + \ell (\sigma(s_{j}))\le L \bigg] \Bigg), \label{eq:F_opt_single}
\end{align}
} where we adapted the expression from~\eqref{eq:F_opt} to the case of a single distribution and a single code. But now trying to extend the recursive expression for $F(k,N)$ in~\eqref{eq:recur_F} gives
\begin{align}
F(k,N) =   &\max \Bigg( \max_{\ell_0 \in [1, L]} \, \bigg( F(k-1, N - N_{\ell_0})
+ p_{k} \cdot  \sum _{i=1}^{n}  p_{i} \cdot I \Big[ \ell_0 + \ell (\sigma(s_{i}) )   \le L \Big] \bigg), F(k-1,N) \Bigg),\label{eq:recur_F_single_extended}
\end{align}
which cannot be used by the algorithm because the indicator function now depends on lengths of codewords that were not assigned yet, namely for the elements $i \in [k+1,n]$. So even though we now only have a single code to design, this task is considerably more challenging than the conditional optimization of Section~\ref{subsec:cond_opt}. At this point the only apparent route to solve~\eqref{eq:F_opt_single} is by trying exhaustively all length assignments to $S$ satisfying Kraft's inequality, and enumerating the arguments of the max function in~\eqref{eq:F_opt_single} directly. But this would be intractable.

\subsection{Efficient algorithm for optimal entry encoding}
In the remainder of the section we show an algorithm that offers an efficient way around the above-mentioned difficulty to assign codeword lengths to elements. We present this efficient algorithm formally, but first note its main idea.\\
\textbf{The main idea}: we showed in Section~\ref{subsec:single_observ} that it is {\em not} possible to maximize the single-code success probability for $k$ elements given the optimal codeword lengths for $k-1$ elements. So it does not work to successively add elements to the solution while maintaining optimality as an invariant. But fortunately, it turns out that it does work to successively add {\em codeword lengths} to the solution while maintaining optimality as an invariant. The subtle part is that the lengths need to be added in a carefully thought-of sequence, which in particular, is {\em not} the linear sequence $(1,2,\ldots,L-1)$ or its reverse-ordered counterpart. We show that if the codeword lengths are added in the order of the sequence
\begin{equation}(L/2,L/2+1,L/2-1,L/2+2,L/2-2,\ldots,L-1,1)\label{eq:len_seq}\end{equation}
(for even\footnote{For convenience we assume that $L$ is even, but all the results extend to odd $L$.} $L$), then for any sub-sequence we can maximize the success probability given the optimal codeword lengths taken from the sub-sequence that is one shorter. For example, when $L=8$ our algorithm will first find an optimal code only using codeword length $L/2=4$; based on this optimum it will find an optimal code with lengths $4$ and $5$, and then continue to add the codeword lengths $3,6,2,7,1$ in that order.

We now turn to a more formal treatment of the algorithm. We first define the function holding the optimal success probabilities for sub-problems of the problem instance. The following Definition~\ref{def:G} is the adaptation of Definition~\ref{def:F_k_N} to the sequence of codeword lengths applicable in the single-code case.
\begin{definition}\label{def:G}
Consider assignments of finite codeword lengths to the consecutive elements $\{s_{\kl},\ldots,s_{\kh}\}$ from $S$, where the lengths are assigned from the values $\{L/2,L/2+1,L/2-1,\ldots,\len\}$ taken from the sub-sequence of~\eqref{eq:len_seq} that ends with $\len$. For $N \in [0, 2^{L}]$ we denote by $\ftag(\len,[\kl,\kh],N)$ the maximal success probability for such an assignment whose sum of weights for these $\kh-\kl+1$ codewords is at most $N$. Formally,
\SingleColMode{
\begin{align}
\ftag(\len,[\kl,\kh],N) =   &\max_{\sigma : \parent{\forall i \in [\kl,\kh]: \ell (\sigma(s_{i}))\in\{L/2,\ldots,\len\}~,~\sum _{i=\kl}^{\kh} {N_{\ell (\sigma(s_{i}))}}  \le N} } \, \Bigg( \sum _{i=\kl}^{\kh} \sum _{i'=\kl}^{\kh} p_{i} \cdot p_{i'}
 \cdot \emph{I} \bigg[ \ell (\sigma(s_{i}) ) + \ell (\sigma(s_{i'}))\le L \bigg] \Bigg). \label{eq:ftag_opt}
\end{align}
}
\end{definition}
The following two theorems are the key drivers of the efficient dynamic-programming algorithm finding the optimal code.
\begin{theorem}\label{th:recur_right}
Let $\len=L/2-\dc+1$ for some integer $1\leq\dc\leq L/2-1$. For the length subsequent to $\len$ in the sequence~\eqref{eq:len_seq} $\len'=L/2+\dc$, we have the following
\begin{align}
\ftag(\len',[\kl,\kh],N) =   \max_{j\in [0,\kh-\kl+1]} \, \ftag(\len,[\kl,\kh-j],N-j\cdot N_{\len'}), \label{eq:ftag_opt_right}
\end{align}
where we define
\begin{equation}\ftag(\len,[\kl,\kl-1],N) \triangleq 0, ~\mathrm{for}~ N\geq 0. \label{eq:extreme_G}\end{equation}
\end{theorem}
\begin{proof}
Given maximal values $\ftag$ for all values of $N$ and with lengths up to $\len$ in the sequence, the maximal value $\ftag$ when $\len'$ is also allowed is obtained by assigning length $\len'$ to between $0$ and $\kh-\kl+1$ elements in the range $[\kl,\kh]$. By the monotonicity of $p_i$, $\len'$ which is {\em higher} than all previous lengths must be assigned to the highest $j$ indices in the range $[\kl,\kh]$. Thus for each $j$ the success probability is the value of $\ftag$ for the corresponding range of elements $[\kl,\kh-j]$ with the residual weight $N-j\cdot N_{\len'}$. In particular, the elements assigned length $\len'$ do not add to the success probability, because $\len'$ plus any length in the sub-sequence up to $\len$ exceeds $L$. In the extreme case when $j=\kh-\kl+1$ (all elements assigned length $\len'$), the definition~\eqref{eq:extreme_G} when appearing in the right-hand side of~\eqref{eq:ftag_opt_right} gives a valid assignment with success probability $0$ if $N$ in the left-hand side is sufficiently large.
\end{proof}

\begin{algorithm}[b!]
	\label{algorithm_optimal_encoding_function_single}
	\SetInd{.03in}{.03in}
	\SetKwInOut{Input}{input}
	\SetKwInOut{Output}{output}
	\Input{Element distribution $(S,P)$, memory width $L$}
	\Output{Prefix code $\sigma$ with optimal entry-encoding success probability}
	
	\caption{Optimal Single Prefix Code}
	
	\underline{{\em initialization}:}\\
	\lForEach{$N < 0$}{
    $\ftag(\len,[\kl,\kh],N) = -\infty$ for all indices $\len$ and $\kl \leq \kh$
    }
	\lForEach{$N \ge 0$}{
		$\ftag(\len,[k,k-1],N) = 0$ for all indices $\len$ and $k$
	}
	
	\underline{{\em codewords of length $L/2$}:}\\


	\ForEach{$[\kl,\kh]\subseteq [1,n]$,  $\kl \leq \kh$}{
    \For{$N=(\kh-\kl+1) \cdot N_{L/2}:2^L$}{
		$\ftag(L/2,[\kl,\kh],N) = \left(\sum_{i=\kl}^{\kh}p_i\right)^2$
         }
         }

\underline{{\em main iteration}:}\\
\For{$\dc=1:L/2-1$}{
          \underline{{\em go right to length $L/2+\dc$}:}\\
          \For{$N=0:2^L$}{
          \ForEach{$[\kl,\kh]\subseteq [1,n]$,  $\kl \leq \kh$}{

            $\ftag(L/2+\dc,[\kl,\kh],N) =   \max_{j\in [0,\kh-\kl+1]} \, \ftag(L/2-\dc+1,[\kl,\kh-j],N-j\cdot N_{L/2+\dc})$\\

	        }
}
\underline{{\em go left to length $L/2-\dc$}:}\\
          \For{$N=0:2^L$}{
          \ForEach{$[\kl,\kh]\subseteq [1,n]$,  $\kl \leq \kh$}{

            $\ftag(L/2-\dc,[\kl,\kh],N) =   \max_{j\in [0,\kh-\kl+1]} \, \left[\ftag(L/2+\dc,[\kl+j,\kh],N-j\cdot N_{L/2-\dc})+\left(\sum_{i=\kl}^{\kl+j-1}p_i\right)\left(\sum_{i=\kl}^{\kh}p_i\right) + \left(\sum_{i=\kl+j}^{\kh}p_i\right)\left(\sum_{i=\kl}^{\kl+j-1}p_i\right) \right] $\\

	        }
}
}		
	\underline{{\em output results}:}\\
	Encoding success probability $P_{success} = \max_{k\in [1,n]} \, \ftag(1,[1,k],2^{L})$\\
	
\end{algorithm}

An important property provided by the length sequence~\eqref{eq:len_seq} is that the success probability up to $\len'$ depends only on the success probability up to $\len$, no matter what the assignments are among the lengths up to $\len$. With this property Theorem~\ref{th:recur_right} allows to efficiently extend the optimality from $\len$ of type $\len=L/2-\dc+1$ (with $\dc \ge 1$) to the next length in the sequence. To complete what is required for an efficient algorithm, we need the same extension of optimality from $\len$ of type $\len=L/2+\dc$ to the next length in the sequence. We do this in the next theorem.
\begin{theorem}\label{th:recur_left}
Let $\len=L/2+\dc$ for some integer $1\leq\dc\leq L/2-1$. For the length subsequent to $\len$ in the sequence~\eqref{eq:len_seq} $\len'=L/2-\dc$, we have the following
\begin{align}
\ftag(\len',[\kl,\kh],N) & = \nonumber\\
  \max_{j\in [0,\kh-\kl+1]} & \, \left[\ftag(\len,[\kl+j,\kh],N-j\cdot N_{\len'})+\left(\sum_{i=\kl}^{\kl+j-1}p_i\right)\left(\sum_{i=\kl}^{\kh}p_i\right) + \left(\sum_{i=\kl+j}^{\kh}p_i\right)\left(\sum_{i=\kl}^{\kl+j-1}p_i\right) \right]. \label{eq:ftag_opt_left}
\end{align}
\end{theorem}
\begin{proof}
Given maximal values $\ftag$ for all values of $N$ and with lengths up to $\len$ in the sequence, the maximal value $\ftag$ when $\len'$ is also allowed is obtained by assigning length $\len'$ to between $0$ and $\kh-\kl+1$ elements in the range $[\kl,\kh]$. By the monotonicity of $p_i$, $\len'$ which is {\em lower} than all previous lengths must be assigned to the lowest $j$ indices in the range $[\kl,\kh]$. Now the success probability has two components: first is the success between pairs of elements assigned lengths up to $\len$ in the sequence, and second is the success between element pairs that involve the new length $\len'$ (recall that in Theorem~\ref{th:recur_right} the second component did not exist, but here it does because $\len'$ is lower than all lengths up to $\len$.) The first of the three terms in the summation of~\eqref{eq:ftag_opt_left} gives the first component, and the latter two terms give the second component. Maximization is done as before by considering all possible $j$ with the residual weight $N-j\cdot N_{\len'}$.
\end{proof}
As in Theorem~\ref{th:recur_right}, the length sequence provides the property that maximization in Theorem~\ref{th:recur_left} only depends on the success probabilities of the previous length sub-sequence, without care to the assignment of lengths among the lengths in that sub-sequence.
In the following Algorithm~\ref{algorithm_optimal_encoding_function_single} we formally present the algorithm for finding an optimal single prefix code, building on Theorems~\ref{th:recur_right} and~\ref{th:recur_left}. For terseness we only track the optimal success probabilities $\ftag$, omitting the more technical task of tracking the optimal assigned lengths, which is required to find the optimal code in a real implementation. The noteworthy parts of Algorithm~\ref{algorithm_optimal_encoding_function_single} are:
\begin{itemize}
\item The initialization of $\ftag$ to $-\infty$ for negative $N$, and to $0$ for non-negative $N$ and empty ranges of elements (according to~\eqref{eq:extreme_G}).
\item Starting the length sequence at $\len=L/2$ and calcualting the success probability when all elements in the range are assigned that length.
\item The main iteration following the progressions in the length sequence using Theorems~\ref{th:recur_right} and~\ref{th:recur_left}.
\item Outputing the optimal success probability for the code parameters as a maximization over all $n$ possible numbers of elements assigned codewords, starting from the first element.

\end{itemize}

\emph{Time Complexity}: 
There are $L-1$ iterations and in each $O(2^{L})$ values are calculated for each of $O(n^2)$ ranges of elements, each by considering $O(n)$ possibilities for the number of elements with the new codeword length. It follows that the time complexity of the algorithm is $O(n^3 \cdot 2^{L}  \cdot L)$, which is polynomial in the size of the input, or $O(n^5 \cdot L)$ (recall that $2^{L}$ is at most quadratic in $n$). It is interesting to compare this complexity as one order of $n$ higher than that of Algorithm~\ref{algorithm_optimal_encoding_function} ($O(n^4 \cdot L)$) for the two-code case. We note that the $n^5$ complexity term is a loose bound, because many of the counted iterations are not exercised in a given run.

As mentioned in the beginning of the section, Algorithm~\ref{algorithm_optimal_encoding_function_single} can be extended to the case where the two fields have {\em different distributions} on the same elements, and a single code with optimal success probability is found. We do not explore this generalization in the paper, but it is as simple as replacing some instances of $p_i$ in the algorithm by $q_i$ corresponding to the distribution of the second field.

\begin{example}
\label{example_shared_code}
Consider the entry distribution
$D= [(S, P), (S, P)]$ with  $(S,P)$ = $((s_{1},\dots,s_{n}),(p_{1},\dots,p_{n}))$ satisfying $(p_{1},\dots,p_{n}) = (0.4,0.4,0.08,0.01,0.01,0.01,0.01,0.01,0.01,0.01,0.01,0.01,0.01,0.01,0.01)$ with $n=15$. The width parameter is $L=6$ and assume that a single code is allowed.  A simple possible coding scheme can assign codewords of length $L / 2 = 3$ to the $2^3=8$ most probable elements $s_{1},\dots,s_{8}$ where codewords are not assigned for the remaining elements $s_{9},\dots,s_{15}$. This scheme successfully encodes all pairs of elements composed of two elements from $s_{1},\dots,s_{8}$ achieving success probability of $P'_{success} = \Big( \sum _{i=1}^{8}  p_{i} \Big)^2 = 0.93^2 = 0.8649$. Alternatively, the above optimal algorithm finds a code that achieves higher success probability.
In this code elements $s_{1},s_{2}$ are assigned codewords of length 2 while elements $s_{3},\dots,s_{10}$ are assigned codewords of length 4. Later elements are not assigned codewords. This optimal scheme achieves an improved success probability of
$P_{success} = \Big( \sum _{i=1}^{2}  p_{i} \Big)^2 +  \sum _{i=1}^{2}  p_{i} \cdot \sum _{i=3}^{10}  p_{i}  +  \sum _{i=3}^{10}  p_{i} \cdot \sum _{i=1}^{2}  p_{i} =
0.8^2 + 0.8 \cdot 0.15 + 0.15 \cdot 0.8 = 0.88$.
\end{example}

\ExtendedJournal{Another example is the lower bound, presented for the case of $d=2$ in Theorem~\ref{theorem_success_probability_lower_bound}.
Let $\ell_1,\ell_2, \cdots, \ell_d$ be positive integers satisfying $\sum _{j=1}^{d} \ell_j = L$.
Any entry satisfying for $j \in [1,d]$ that its $j^\text{th}$ field is encoded within $\ell_j$ bits is encoded successfully. Accordingly, a lower bound on the success probability that equals
$\Big( ( \sum _{i=1}^{2^{\ell_1}-1}  p_{1, i}) \cdot ( \sum _{i=1}^{2^{\ell_2}-1}  p_{2, i}) \cdot \cdots \cdot   ( \sum _{i=1}^{2^{\ell_d}-1}  p_{d, i}) \Big)$
can be easily deduced. Also in the case of $d > 2$ fields, there exists a monotone \st{en}coding scheme that is optimal. Accordingly, the number of solutions that have to be examined in order to find an optimal \st{en}coding scheme is given by $Z_{n_1} \cdot Z_{n_2} \cdot \cdots \cdot Z_{n_d}$.
Likewise, the joint \ArcName architecture can be easily generalized to support more than two fields by simply increasing the number of dictionaries.
In Section~\ref{Section_Experimental_Results}, we describe an experiment of encoding entries with more than two fields.}


\section{Experimental Results}
\label{Section_Experimental_Results}

We examine the performance of the suggested coding schemes.  
In the experiments, the probabilities in the element distributions follow the Zipf distribution with different parameters.
A low positive Zipf parameter $\mu$ results in a distribution that is close to the uniform
distribution, while for a larger parameter the distribution is more biased.

\begin{figure}[!b]
\subfigure[two distributions, $\mu_1 = 0.8, \mu_2=2$] {
\includegraphics[width= 0.44 \textwidth]{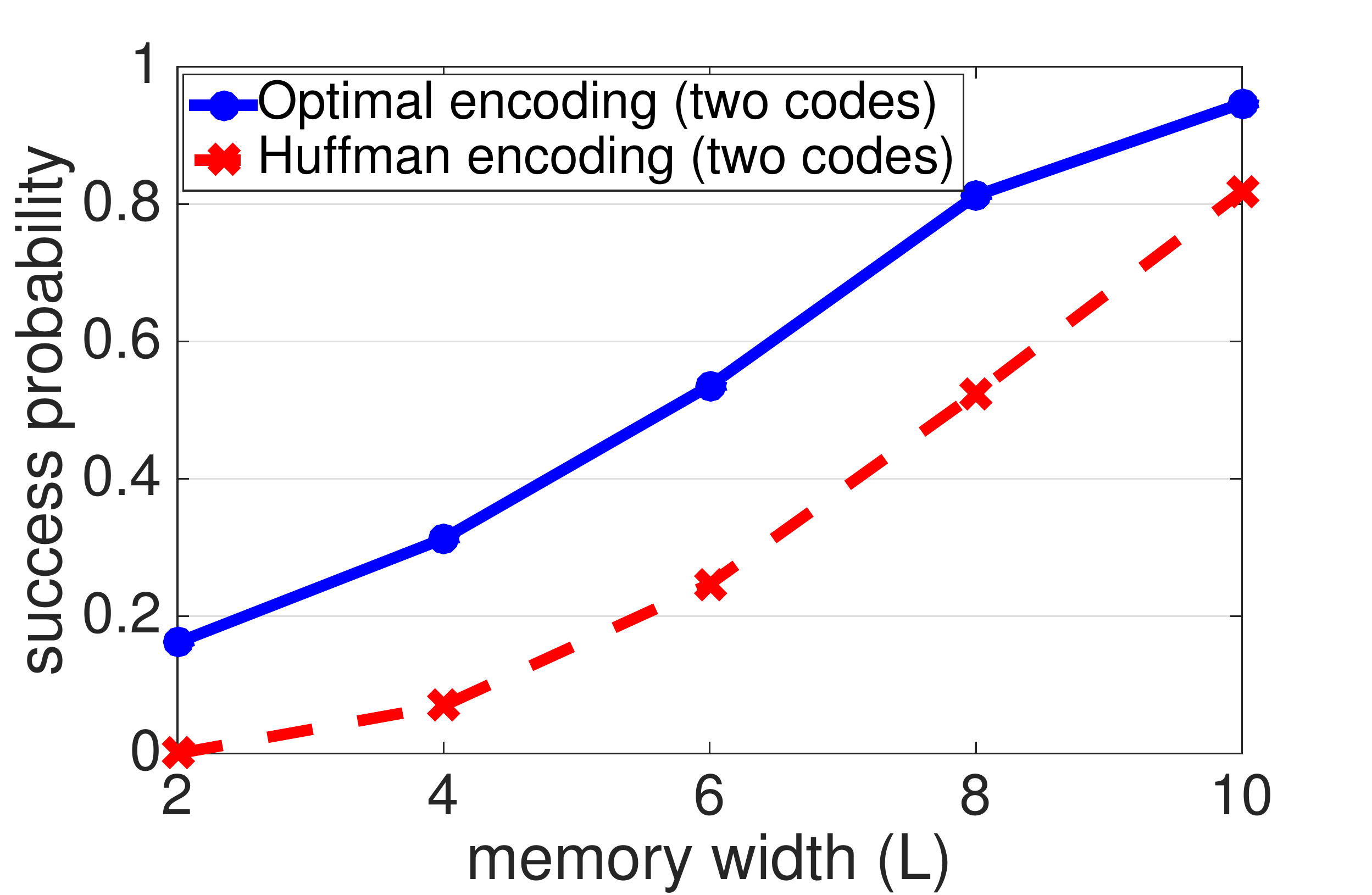}
\label{fig:experiments_Huffman_comparison_a}}
\hfill
\subfigure[single distribution, $\mu = 1.6$] {
\includegraphics[width=0.44  \textwidth]{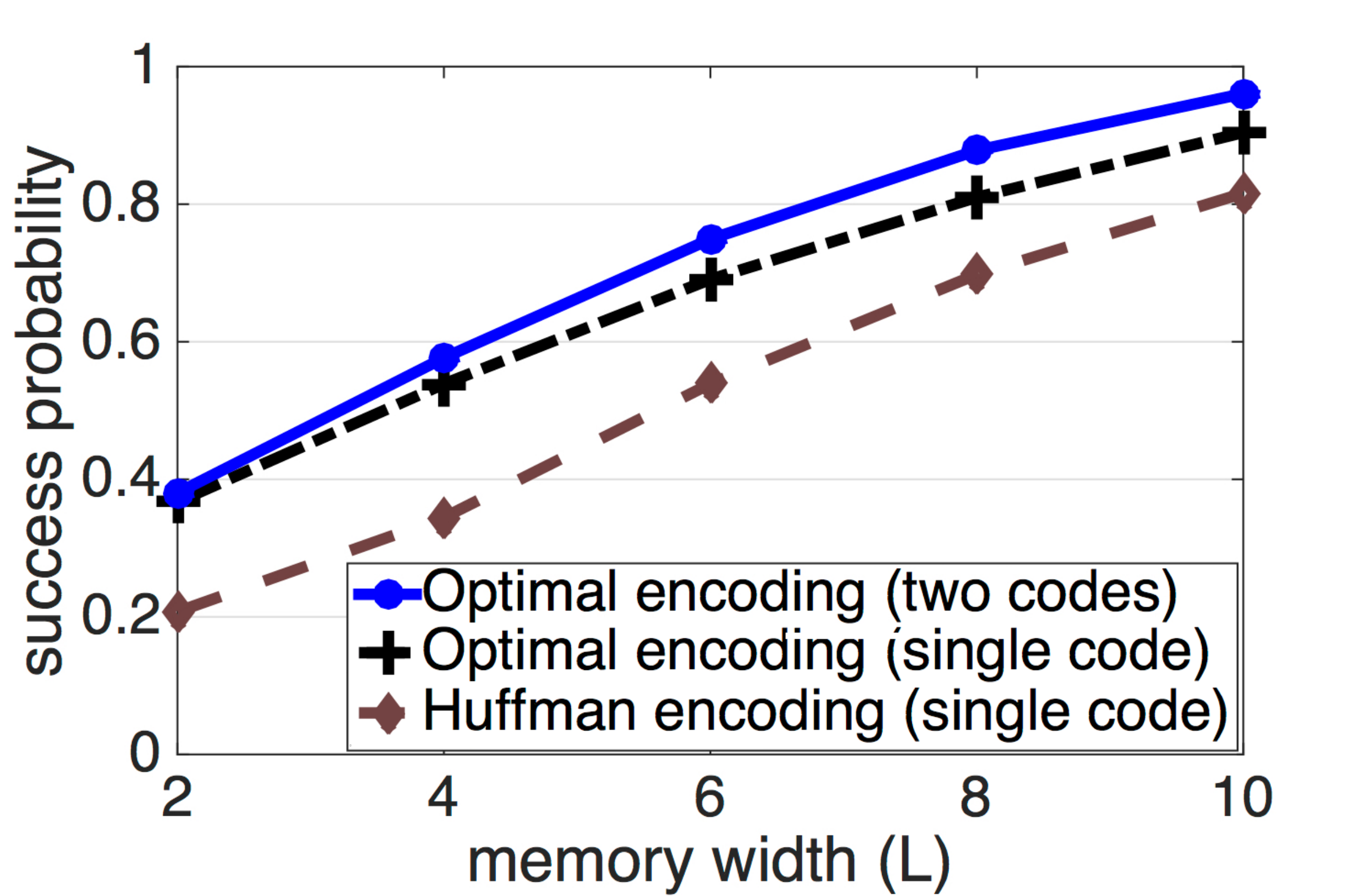}
\label{fig:experiments_Huffman_comparison_b}}
\caption{
Comparison of the optimal coding schemes for two codes and a single shared code vs. Huffman-based schemes.
In (a), with two codes for fields with two distributions. The optimal coding scheme (from Section~\ref{sec:optimal_coding}) is compared with a scheme composed of the two Huffman codes for the two fields.
In (b), with a single shared code for field with the same distribution. The optimal coding scheme (from Section~\ref{Section_single_distribution}) is compared with a scheme composed of the same Huffman code for the two fields.
Probabilities follow Zipf distribution with parameters  $\mu_1=0.8, \mu_2=2$ (in (a)) and $\mu=1.6$ (in (b)).
}\label{fig:experiments_Huffman_comparison}
\end{figure}

\begin{figure}[!t]
\subfigure[single code, $\mu = 0.5$] {
\includegraphics[width= 0.44 \textwidth]{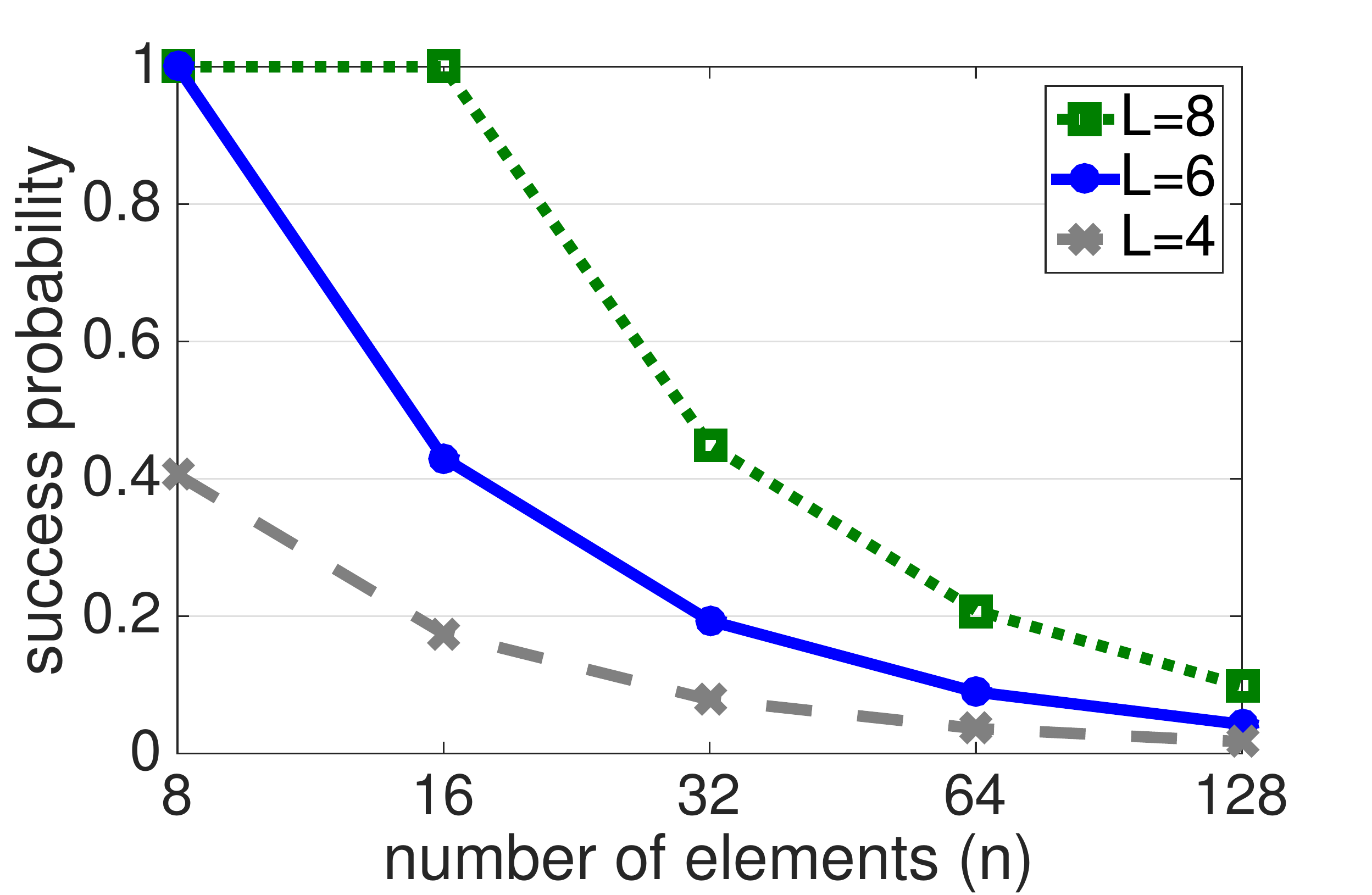}
\label{fig:experiments_success_vs_n_a}}
\hfill
\subfigure[single code, $\mu = 2$] {
\includegraphics[width=0.44  \textwidth]{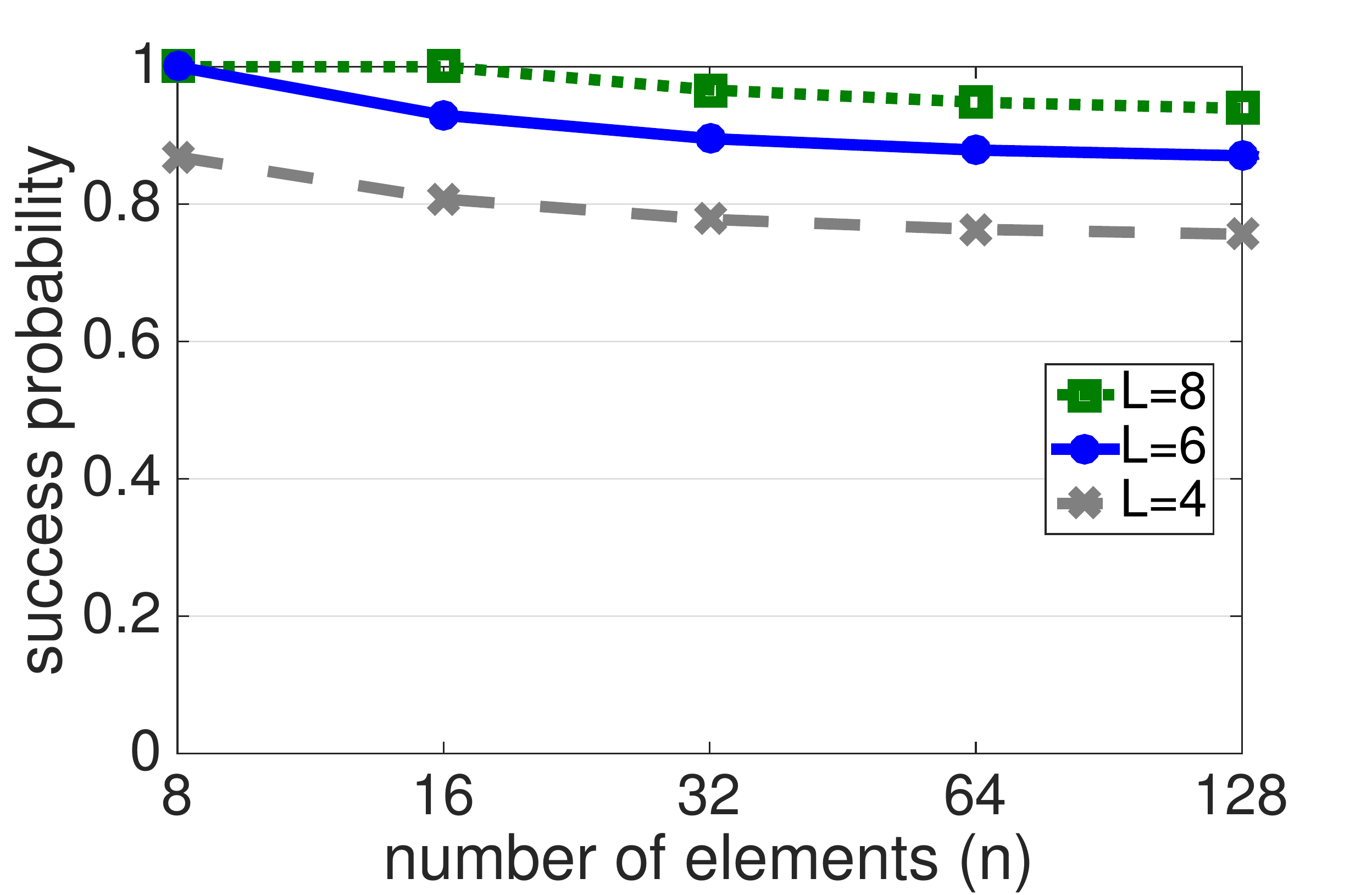}
\label{fig:experiments_success_vs_n_b}}
\caption{
The success probability of the optimal coding scheme for a shared single code vs. the number of elements in the distribution for different values of the memory width $L$.
Probabilities follow the Zipf distribution with parameters  $\mu=0.5$ (in (a)) and $\mu=2$ (in (b)).
}\label{fig:experiments_success_vs_n}
\end{figure}

As a first step, we compare the suggested optimal coding schemes
to Huffman coding~\cite{Huffman}. The results are shown in Fig.~\ref{fig:experiments_Huffman_comparison}.
First, in Fig.~\ref{fig:experiments_Huffman_comparison_a}, we pick two different distributions on 128 elements for the two fields, and  set the assumption that the codes of the two fields may be different. The  distributions of the first and second field are Zipf with parameters $\mu_1=0.8$ and $\mu_2=2$, respectively. We compute the optimal codes $\sigma_1$ and $\sigma_2$ using Algorithm~\ref{algorithm_optimal_encoding_function} from Section~\ref{sec:optimal_coding}, and plot their success probability. In comparison, we also plot the success probability of the Huffman code computed for each field according to its distribution. It can be seen that while Huffman codes minimize the expected codeword length, in a fixed-width memory their encoding success probability is inferior to the optimal code for all values of $L$.
For $L=2$ for instance,  Huffman coding does not encode successfully any pair of elements, while the optimal coding scheme achieves a probability of 0.162.
A maximal difference in the success probability of 0.289 is obtained for $L=6$.

In Fig.~\ref{fig:experiments_Huffman_comparison_b} we examine the scenario of the two fields having the same element distribution, with the requirement that a single code is shared by the two fields. An optimal coding scheme for this case was given in  Section~\ref{Section_single_distribution}.  The two fields have the same 128 elements with a Zipf $\mu = 1.6$ distribution. We compute the optimal code $\sigma$ using Algorithm~\ref{algorithm_optimal_encoding_function_single}, and plot its success probability. In comparison, we also plot the success probability of the Huffman code computed from the common element distribution. Here too Huffman coding gives inferior success to the optimal code, with a maximal probability difference of 0.194 at $L=4$. We also plot for comparison the success probability of the optimal coding scheme when allowing two different codes for the two fields. The results underscore the fact that allowing two different codes can improve success probability even when the two fields follow the same distribution.

Next, we examine the impact of the number of possible elements in the fields for different Zipf distributions. We assume a single shared code and that the two fields follow  the same distribution. In Fig~\ref{fig:experiments_success_vs_n}, we consider two
such distributions and present the optimal success probability.
In  Fig~\ref{fig:experiments_success_vs_n_a}, the distribution has  a  parameter $\mu=0.5$. We assume that there are $n$ elements in each field.
The minimal width required to obtain a success probability of 1 for $n$ elements is given by $L = 2 \cdot \log_2 n$, i.e. $L \ge 6$ for $n=8$ and $L \ge 8$ for $n=16$. For a given $L$, the optimal success probability  decreases when the number of elements in each field increases. For $L=8$ it equals 0.449, 0.208 and 0.099 for $n=32,64$ and $128$, respectively.
Likewise, in  Fig~\ref{fig:experiments_success_vs_n_b}, the distribution has  a  parameter $\mu=2$, which results in a more biased element distribution. While again for $n=8$, a success probability of 1 is observed only for $L \ge 6  = 2 \cdot \log_2 n$, the more  biased distribution allows achieving high success probabilities even for larger values of $n$. For instance, while for $n=128$ a memory width of $L = 2 \cdot \log_2 n = 14$ is required to guarantee a success probability of 1, already for $L=8$ we obtain probability of 0.939 when $\mu=2$.

\section{Conclusion}
\label{Section_conclusions}
In this paper we developed compression algorithms for fixed-width memories. We presented a new optimization problem of maximizing the probability to encode entries within the memory width, and described constraints on the codes for the different input fields that guarantee uniquely-decodable data entries.
We derived algorithms that find an optimal coding scheme with maximum success probability for the scenario of two codes for two fields, as well as for a single shared code. We also presented experimental results to evaluate the success-probability performance.
The central open problem left by this paper is extending optimal coding to $d>2$ fields.

\section{Acknowledgment}
We would like to thank Neri Merhav for fruitful discussions and pointers.


\appendix[Analyzing the number of monotone prefix codes]

To demonstrate the hardness of the problem of finding an optimal coding scheme we discuss the number of possible solutions that have to be considered.
Consider an entry distribution $D= [(S_1, P_1), (S_2, P_2)]$ = $\Big[((s_{1, 1},\dots,s_{1, n_{1}})$,$(p_{1, 1},\dots,p_{1, n_{1}})),$ $((s_{2, 1},\dots,s_{2, n_{2}})$,$(p_{2, 1},\dots,p_{2, n_{2}}))\Big]$ (with $n_1 = |S_1|$, $n_2 = |S_2|$).

We recall that a solution for the  problem of Section~\ref{sec:optimal_coding} is composed of a prefix code as well as a \pinv code. Likewise, a solution for the  problem of Section~\ref{Section_single_distribution} is a prefix code.
We show that even when considering only monotone coding schemes (as pointed out by Lemma~\ref{lemma_codewords_lengths_order}), i.e., codes for which longer codewords are assigned to less probable elements, the number of such prefix codes grows exponentially as a function of the element counts $n_j$.
We identify a prefix code for $S_j$ by a
non-decreasing vector of $n_j$ codeword lengths. We recall that the codeword lengths are sufficient to determine the success probability, even without knowing the actual codewords.

Finding the optimal code can be reduced to examining binary trees of a special kind.
The number of leaves in each tree is equal to $n$, the number of codewords in the code.
The depth of each of the $n$ leaves in the tree represents the length of the
corresponding codeword  (the depth of a leaf is the length of the path from the root to the leaf).
We also assume that the set of depths (and accordingly the codewords lengths) attain Kraft's inequality with equality. Otherwise, some codewords can be shortened.
To satisfy the requirement on the non-decreasing order of the lengths, we consider only trees with the property that the depth of a leaf equals at least the depths of all the leaves left of it in the tree. We say that a tree is monotone if it satisfies this property.

\begin{figure}[!t]
\centering
\subfigure[$n=2, Z_2 = 1$: a single possible vector $(1,1)$.] {
\includegraphics[width=0.12 \textwidth]{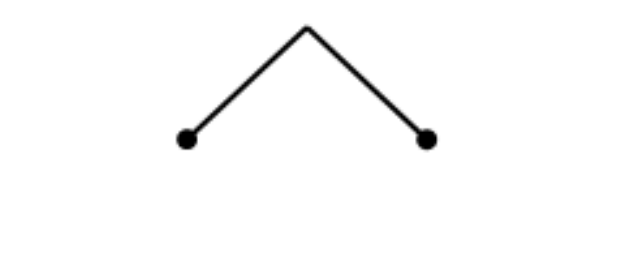}
\label{fig:tree_2}}
\hfill
\subfigure[$n=3, Z_3 = 1$: a single possible vector $(1,2,2)$.] {
\includegraphics[width=0.12 \textwidth]{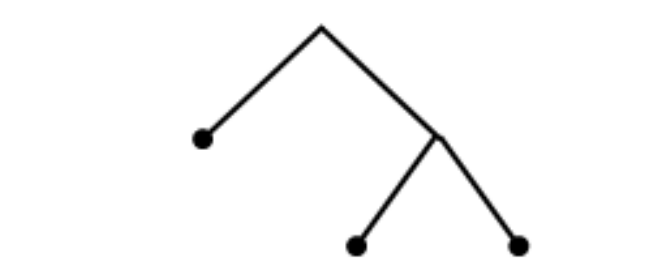}
\label{fig:tree_3}}
\hfill
\subfigure[$n=4, Z_4 = 2$: Two vectors: $(1,2,3,3)$ and $(2,2,2,2)$.] {
\includegraphics[width=0.33 \textwidth]{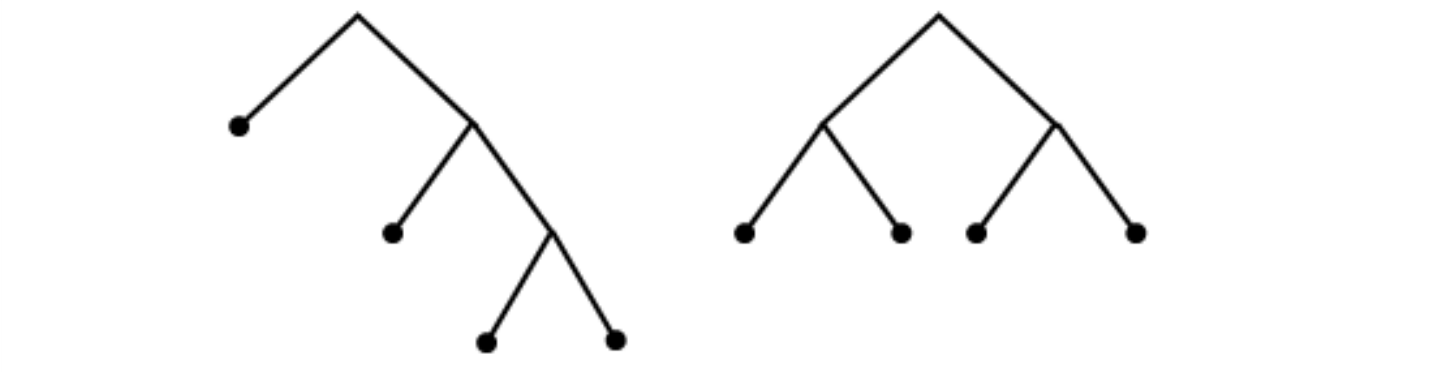}
\label{fig:tree_4}}
\hfill
\subfigure[$n=5, Z_5 = 3$: Three vectors: $(1,2,3,4,4)$, $(1,3,3,3,3)$ and $(2,2,2,3,3)$.] {
\includegraphics[width=0.33 \textwidth]{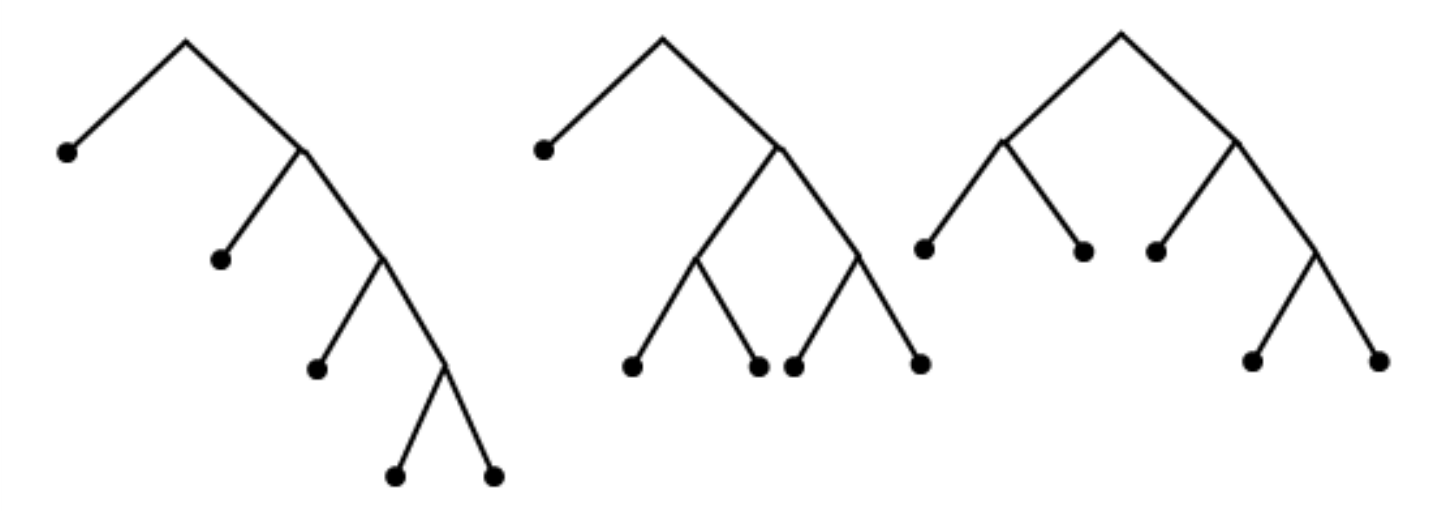}
\label{fig:tree_5}}
\subfigure[$n=6, Z_6 = 5$: Five vectors: $(1,2,3,4,5,5)$, $(1,2,4,4,4,4)$, $(1,3,3,3,4,4)$, $(2,2,2,3,4,4)$ and $(2,2,3,3,3,3)$.] {
\includegraphics[width=0.8 \textwidth]{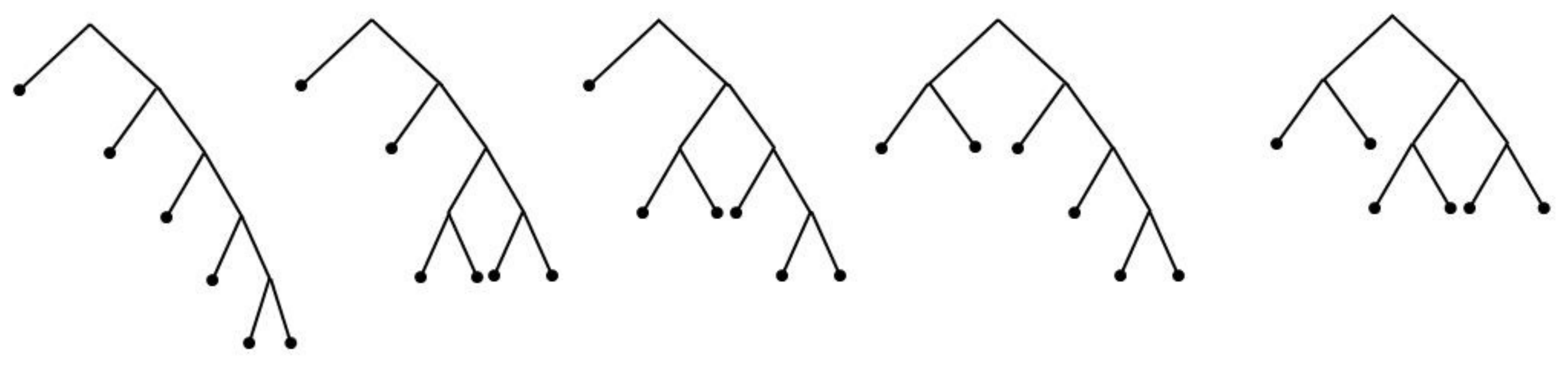}
\label{fig:tree_6}}
\caption{Illustration of the binary trees representing the possible codes of $n$ elements (for $n \in [2,6]$).
Each code is characterized by a vector of $n$ codeword lengths. For each value of $n$, the represented vectors are ordered by the lexicographic order.
}\label{fig:tree}
\end{figure}

\begin{example}\label{exp_tree_representation}
Fig.~\ref{fig:tree} presents for $n \in [2,6]$ the  monotone binary trees for the possible codes  of  $n$ elements with their corresponding length vectors. For each value of $n$, the represented vectors are ordered by the lexicographic order.
In each tree, the depth of the $i^\text{th}$ leaf (in left to right order) equals the $i^\text{th}$ length in the vector. For instance, in Fig.~\ref{fig:tree_4} (for the case of $n=4$), we can see two trees that represent the two possible vectors $(1,2,3,3)$, $(2,2,2,2)$.
Accordingly, in a prefix code of four elements, we can have one option of one codeword of a single bit, another codeword of two bits and two additional codewords of three bits. Alternatively, we can have four codewords of two bits.
Likewise, for $n=5$ there are 3 possible vectors
$(1,2,3,4,4)$, $(1,3,3,3,3)$ and $(2,2,2,3,3)$ (shown in Fig.~\ref{fig:tree_5}) and for $n=6$ there are $5$ vectors $(1,2,3,4,5,5)$, $(1,2,4,4,4,4)$, $(1,3,3,3,4,4)$, $(2,2,2,3,4,4)$ and $(2,2,3,3,3,3)$  (Fig.~\ref{fig:tree_6}).
\end{example}

Toward counting the number of monotone trees, we can observe that a monotone tree with $n$ leaves is composed of two monotone subtrees with $j$ and $(n-j)$ leaves for some $j, n-j$. In addition, in order to keep the full tree to be monotone,
the depth of the leftmost leaf in the right subtree must equal at least the depth of the rightmost leaf in the left subtree. This constraint yields that $j \le (n-j)$, i.e. the number of leaves in the left subtree cannot be larger than the number in the right subtree. To search for tree of $n$ leaves, representing a code of $n$ elements, we simply consider all options to build it from two trees of $j$ and $(n-j)$ leaves satisfying the above property for their depths.

The next example shows how we can deduce the possible monotone trees for $n=5$, $n=6$ based on the possible trees for smaller values of $n$.
\begin{example}\label{exp_alternative_approach}
We first show how to calculate for $n=5$ the monotone trees and the corresponding lengths vectors by relying on the trees for $n \in [1,4]$.
We explained that if $n=1$ there is a single tree with one node of depth (0). As presented in Fig.~\ref{fig:tree}, for $n=2$ there is a single tree with depths of $(1,1)$ and for $n=3$ a single tree with depths of $(1,2,2)$. For $n=4$ the two possible vectors are $(1,2,3,3)$ and $(2,2,2,2)$. For $n=5$, we consider the values of $j \in [1,\lfloor \frac{n}{2} \rfloor] = [1,\lfloor \frac{5}{2} \rfloor]= [1,2]$ as a possible number of leaves in the left subtree.
If $j=1$, the tree represented by the vector $(0)$ (of $j=1$ element) can be combined with any of the two trees with $(n-j)=(5-1)=4$ nodes represented by the vectors $(1,2,3,3)$ and $(2,2,2,2)$. This results in two possible trees of $n=5$ elements with depths of $(0+1,1+1,2+1,3+1,3+1), (0+1,2+1,2+1,2+1,2+1) = (1,2,3,4,4), (1,3,3,3,3)$. We also consider the case $j=2$. Here, the tree of the vector $(1,1)$ (of $j$ elements) can be combined with the tree of the vector $(1,2,2)$ (of $(n-j)=3$ elements) since $1 \le 1$. Accordingly, we have an additional option for a tree of $n=5$ elements with depths  $(1+1,1+1,1+1,2+1,2+1) = (2,2,2,3,3)$. To conclude, we have three trees representing three codes of $n=5$ elements with codewords of lengths $(1,2,3,4,4),(1,3,3,3,3)$ or $(2,2,2,3,3)$. In a similar way, we can calculate the possible trees for $n=6$. The single tree for $j=1$ can be combined with any of the three trees for $n-j=5$. The single tree for $j=2$ can be combined with any of the two trees for $n-j=4$ while the single tree  for $j=3$ with depths of $(1,2,2)$ cannot be combined with itself (here, $n-j=3=j$) since $1 < 2$. Therefore, we have a total number of $3+2=5$ possible trees of $n=6$ elements.
\end{example}

A simple property of a length vector of $n$ elements was described in~\cite{BooksDaglib0016881}.
\begin{property}\label{property_possible_lengths_vectors}
Let $(\ell_1,\dots,\ell_n)$ be a length vector of $n$ elements, i.e., it satisfies $\ell_{i_1} \le \ell_{i_2}$ if $i_1 < i_2$,  $\ell_{j} \in \mathbb{N^+}$ and $\sum _{j=1}^{n} {2^{- \ell_{j}} } = 1$. Then, $\ell_{n} \le n-1$ and
$\ell_{n-1} = \ell_{n}$.
\end{property}

We denote by $Z_n$ the number of possible monotone trees with $n$ codeword lengths  $(\ell_1,\dots,\ell_n)$.
As mentioned, they satisfy $Z_1 = Z_2 = Z_3 = 1$, $Z_4 = 2$, $Z_5 = 3$ and $Z_6 = 5$.
We  study the function $Z_n$ in order to bound the complexity of examining all the code trees to find the optimal one exhaustively.

Consider a monotone tree of $n$ elements. For $y > 1$, we can obtain a monotone tree of $n + y$ elements by replacing the most right node of the tree by a monotone tree of size $y+1$.  Notice that the modified tree of size $n + y$ must also be monotone since the depth of the new nodes must be larger than the depth of the eliminated node. We can see that starting from some fixed $n$-element tree, using distinct trees of size $y+1$ results in distinct trees of size $n+y$. Likewise, two distinct trees of size $n$ must differ in the depth of some of their first $n-1$ elements,  thus they always yield different trees of size $n+y$ after any possible replacements.
We recall that $Z_4 = 2$, i.e., there are two monotone trees of size 4. We can derive $2 \cdot Z_n$ distinct trees of size $n+3$ by applying the replacement on each of the $Z_n$ trees of size $n$ using each of the   $Z_4 = 2$ trees of size 4. Thus necessarily  $Z_{n+3} \ge 2 \cdot Z_n$.
More generally, we can show using similar possible replacements that $Z_{n+\alpha} \ge Z_{\alpha + 1} \cdot Z_n$
for $n \ge 1, \alpha \ge 1$. This yields that $Z_{n+3} \ge 2 \cdot Z_n$, $Z_{n+4} \ge 3 \cdot Z_n$, $Z_{n+5} \ge 5 \cdot Z_n$, $Z_{n+6} \ge 9 \cdot Z_n$ etc.

We  present a recursive formula for $Z_n$.
To do so, we define for a given $n$ by $A_{i,\ell}^{x}$ the number of non-decreasing vectors of $i$ integer elements of the form
$(\ell_1,\dots,\ell_i)$ satisfying $\ell_j \ge \ell\, \,  \forall j \in [1,i]$ and $\sum _{j=1}^{i} {2^{- \ell_{j}} } = x$.
By definition, we have that $Z_n$ is given by $A_{i,\ell}^{x}$ for $i=n$, $\ell=1$ and $x=1$.
For the correctness of the following recursive formulas, we set $A_{1,\ell}^{x}=1$ for $x \in [2^{-\ell}, 2^{-(\ell+1)}, \dots, 2^{-(n-1)}]$ (for each value of $x$, only the vector with a single length of $-\log_2(x) \ge \ell$ is possible) and  $A_{i,\ell}^{x}=0$ for
$i \ge 1, x \le 0$.
We show the following property.
\begin{property}\label{property_recursive_formula_A}
For a given $n$, the function $A_{i,\ell}^{x}$ satisfies
\begin{align}
A_{i,\ell}^{x} =  \sum _{r=\ell}^{n-1}  A_{i-1,r}^{x-2^{-r}}.
\end{align}
In particular,
\begin{align}
Z_n = A_{i=n,\ell=1}^{x=1} =  \sum _{r=1}^{n-1}  A_{n-1,r}^{1-2^{-r}}.
\end{align}
\end{property}
\bp
We deduce from Property~\ref{property_possible_lengths_vectors} an upper bound on the maximal codeword length.
For the vector of $i$ lengths $(\ell_1,\dots,\ell_i)$,  we consider all possible options for the value of $\ell_1 \in [\ell,n-1]$. By denoting this value by $r$, we must have that the last $i-1$ lengths satisfy $\sum _{j=2}^{i} {2^{- \ell_{j}} } = x-2^{-r}$. Since the vector is non-decreasing, they also have a value of at least $r$. Clearly, all these cases are disjoint. The result for $Z_n$ directly follows the explanation from above.
\ep

We derive a lower bound for $Z_n$. From the inequality $Z_{n+3} \ge 2 \cdot Z_n$ and the values $Z_4 = 2$, $Z_5 = 3$ and $Z_6 = 5$ we have for instance that $Z_n \ge 2 \cdot 2$ for $n \in [7,9]$, $Z_n \ge 2 \cdot 2^2$ for $n \in [10,12]$, $Z_n \ge 2 \cdot 2^3$ for $n \in [13,15]$.
More generally, for $n \ge 4$ $Z_n \ge 2 \cdot 2^{\left \lfloor \frac{n-4}{3} \right \rfloor}$. Thus, for $n \ge 4$
\begin{align}
 Z_n \ge 2 \cdot 2^{\left \lfloor \frac{n-4}{3} \right \rfloor} \ge  2 \cdot 2^ {\frac{n-6}{3}}  = 0.5 \cdot 2^{\frac{n}{3}}  = 0.5 \cdot (\sqrt[3]2)^n.
\end{align}

\begin{figure*}[!t]
\centering
\includegraphics[width= 0.60  \textwidth]{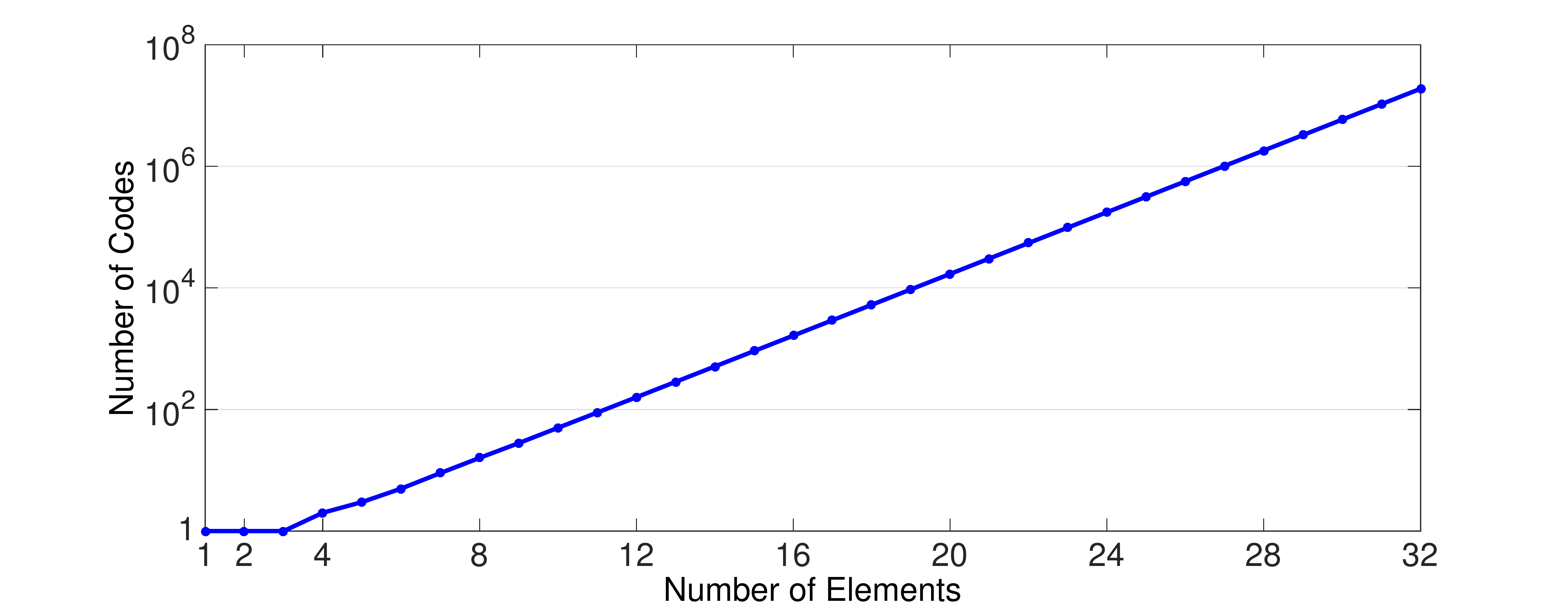}
\caption{The number of codes that have to be considered for a given number of elements in a field.}
\label{fig:solutions_number}
\end{figure*}

Fig.~\ref{fig:solutions_number} shows the value of $Z_n$ for $n \in [1,32]$ in logarithmic scale and demonstrates its exponential increase.

\end{document}